\documentclass[12pt,preprint]{aastex}

\shorttitle{SUMSS GIANT RADIO SOURCES}
\shortauthors{L. Saripalli et. al.}

\begin{document}

\title{A COMPLETE SAMPLE OF MEGAPARSEC-SIZE DOUBLE RADIO SOURCES FROM SUMSS}

\author{L.~Saripalli,\altaffilmark{1} R.~W.~Hunstead,\altaffilmark{2}  
R.~Subrahmanyan\altaffilmark{1} and E.~Boyce,\altaffilmark{2,3}}

\altaffiltext{1}{CSIRO Australia Telescope National Facility, 
Locked bag 194, Narrabri NSW 2390, Australia. Lakshmi.Saripalli@csiro.au and Ravi.Subrahmanyan@
csiro.au}
\altaffiltext{2}{School of Physics, University of Sydney, NSW 2006, Australia.
rwh@physics.usyd.edu.au}
\altaffiltext{3}{MIT Kavli Institute, 77 Massachusetts Avenue, Cambridge MA 02139. 
eboyce@mit.edu}

\begin{abstract}
We present a complete sample of megaparsec-size double radio sources
compiled from the Sydney University Molonglo Sky Survey (SUMSS).  
Almost complete redshift information has been obtained for the sample. The
sample has the following defining criteria: Galactic latitude $\left\vert b \right\vert > 12\fdg5$,
declination $\delta < -50\degr$ and angular size $ > 5\arcmin$. All the 
sources have projected linear size larger than 0.7~Mpc (assuming $H_{o}$ = 
71~km~s$^{-1}$~Mpc$^{-1}$). The sample is chosen from a region of the sky
covering 2100 square degrees. In this paper, we present 843-MHz 
radio images of the extended radio morphologies made using the
Molonglo Observatory Synthesis Telescope (MOST), higher resolution radio observations 
of any compact radio structures using the Australia Telescope Compact Array (ATCA),
and low resolution optical spectra of the host galaxies from the 2.3-m Australian
National University (ANU) telescope at Siding Spring Observatory. 
The sample presented here is the first in the southern hemisphere and 
significantly enhances the database of known giant radio sources. The giant radio sources 
with linear size exceeding 0.7~Mpc have an abundance of (215~Mpc)$^{-3}$ at the sensitivity
of the survey. In the low redshift universe, the survey may be suggesting the 
possibility that giant radio sources with relict lobes are more numerous 
than giant sources in which beams from the centre currently energize 
the lobes.

\end{abstract}
\keywords{ Surveys -- samples: galaxies -- radio galaxies: continuum -- optical spectra}

\section{Introduction}

Double radio sources of large linear size---the giant radio sources---form 
a valuable resource for understanding the radio galaxy phenomenon 
and probing the intergalactic medium \citep{Su93}. Their 
large sizes raise several interesting questions, not only concerning
their formation but also issues related to the long lifetime of their nuclear 
beam activity and jet stability. 
The long timescales involved also imply that the radio structures
of these giant radio sources might show evidence of
changes in nuclear activity that might be linked to recent merger 
events or other triggers of central engine activity.
Recent discoveries of radio sources with inner double structures embedded 
within more relaxed outer diffuse lobes \citep{Su96,Sc00a,Sa02,Sa03}
have drawn attention to the
phenomenon of restarting in beams.
Investigations into these issues clearly benefit from having 
large numbers of giant radio sources and, importantly, samples with 
uniform selection criteria. 

The number of known giant radio sources has grown largely 
from serendipitous discoveries over the years. However, there have
been several attempts in the last decade to make directed searches for 
these large objects. Low-frequency radio surveys with high surface brightness
sensitivity are specially suited for the purpose of detecting giant radio sources.
\citet{Co96} presented a sample of giant radio sources
selected from the 7C survey.  The Westerbork Northern Sky Survey (WENSS; \citet{Ren97}) was used 
by \citet{Sc01} to compile a complete flux-density limited sample 
of giant radio sources. \citet{La01} and also 
\citet{Ma01} compiled samples of large angular-size sources
using the NRAO Very Large Array sky survey (NVSS; \citet{Con98}) and the Faint Images of the Radio
Sky at Twenty cm (FIRST; \citet{Bec95}) and worked towards identifying  
giant radio sources in these surveys.  In this paper we describe a similar effort that
has resulted in the formation of a complete sample of giant radio sources in the southern 
hemisphere based on the Sydney University Molonglo Sky Survey (SUMSS). 

The SUMSS has been described in detail elsewhere \citep{Bo99,Sa01}. 
The 843-MHz SUMSS was started in 1997 and uses 
the upgraded Molonglo Observatory Synthesis Telescope (MOST) as the survey 
instrument. With a synthesized beam of FWHM $45$cosec$\left\vert \delta \right\vert \times 45 $~arcsec$^{2}$ 
at a position angle $0\degr$, an rms noise level of $\sim 1$~mJy~beam$^{-1}$ 
and excellent spatial frequency coverage, 
the survey is one of the most sensitive to date for extended sources.  The sample presented in 
this paper is from the region observed up to the end of February 2000
and covers 2100 square degrees. In Section~2 we describe the
methodology adopted for initially selecting candidate giant radio sources in this
sky region.  Subsequently, we carried out high resolution Australia Telescope 
Compact Array (ATCA) radio observations 
for detecting radio cores in these candidates and this is reported in Section~3. 
Spectroscopy of the optical identifications made at the locations of the radio cores
is presented in Section~4. The resulting giant radio source sample is presented along 
with notes on individual sources in Section~5. In Section~6 we discuss the sample
properties.  The candidates that were rejected while forming the complete sample are
discussed in Appendix~A.

Giant radio sources are usually defined in the literature 
to be double radio sources with projected
linear size greater than 1~Mpc; this limit was based on a Hubble 
constant of $H_{o}$ = 50~km~s$^{-1}$~Mpc$^{-1}$.  
We adopt a flat cosmology with Hubble constant 
$H_{o}$ = 71~km~s$^{-1}$~Mpc$^{-1}$ and matter density parameter 
$\Omega_{m}$ = 0.27.  
Double radio sources with linear 
size exceeding $0.7h_{71}^{-1}$~Mpc are large enough to be atypical and most of
the extended radio structure would lie well outside of gas associated with the host galaxy;
however, there is no reason for adopting a definitive linear-size cutoff 
apart from having well defined sample selection criteria.  
For consistency with previous work, we have chosen to 
include in our final sample all radio galaxies with linear sizes larger than 
$0.7$~Mpc for our adopted $H_{o}$.

\section{The SUMSS giant radio source candidates}

\citet{Bo00} examined the sky area covered by SUMSS prior to 2000 February 
for candidate giant radio sources; at that time SUMSS was 26\% complete.
Boyce also limited the search to 
declinations $\delta < -50\degr$ and Galactic latitudes $\left\vert b \right\vert > 12\fdg5$.  A total of 2100 
square degrees of sky area was examined by eye.  At the low frequency of 843 MHz 
and the low resolution of $\approx 45 \arcsec$, the lobes rather than cores or hotspots 
of extragalactic double radio sources dominate the images. 
A search was made for single connected structures that have a large angular size, close
pairs of extended radio components---double sources that appeared to be
extended along the line joining the two components---and any triple sources. 
The angular extent was estimated to be the separation between the peaks of 
the outermost components and only sources with angular 
size exceeding $5\arcmin$ were deemed to be candidate giant radio sources.  
A total of 35 candidates were selected as potential giant radio sources.

In this paper we adopt a nomenclature for sources that identifies 
the survey and their giant nature. The giant radio 
sources that satisfied the selection criteria are given the 
prefix: SGRS (SUMSS giant radio source), and the 
remainder of the candidate list are given the prefix: SGRSC (SUMSS giant radio source candidate).
It may be noted that some of the candidates in the Appendix have linear sizes exceeding 0.7~Mpc
and are giant radio sources, some are double radio sources with smaller linear size 
and some candidates are composed of unrelated components.

Twelve extended double radio sources  (SGRSC~J0020$-$7321, J0129$-$6433, J0200$-$6007, 
J0534$-$8203, J0551$-$5655, J0603$-$5429, J0622$-$5938, J1959$-$6402,
J2150$-$6210, J2222$-$5617, J2228$-$5600 and J2253$-$5813) were 
detected in the search area with angular sizes, measured from the SUMSS images, that were close to the limit of
$5\arcmin$ and below the angular-size cutoff. 
We included them in follow-up high-resolution radio imaging
to test whether there were any giant radio sources among them. 
As discussed below, the follow-up observations revealed that
some of these double/triple components were unrelated sources
and some had linear sizes smaller than 0.7~Mpc. 

For the two sources SGRC~J0143$-$5431 and J0326$-$7730 the emission peak in one of the lobes
was displaced well away from the edge, unlike the other candidates. For these two sources
we identified components representing the extended emission with a Gaussian fit to a slice 
along the lobe. The angular extent for the sources were estimated using these identified 
components.

The selection criteria included a declination limit so that the 
SUMSS radio images and the follow-up with the E-W ATCA
had synthesized beams that were not too elongated. Low Galactic latitudes were avoided
to facilitate optical identifications.
The lower limit of $5\arcmin$ on the angular size was chosen to include essentially all 
powerful edge-brightened double radio sources 
in the nearby universe with total linear size larger than 0.7~Mpc 
and to ensure that the candidate list did not overwhelm
our resources for radio/optical
follow-up; this criterion was the same as that adopted by \citet{Sc01}.
The angular-size cutoff, together with the sensitivity limit of the survey,
potentially discriminates against edge-darkened (FR-{\sc i} type) double radio sources; 
even if the largest angular size in edge-darkened sources exceeds $5\arcmin$,
it is possible that the separation between the peaks in their lobes 
is less than this cutoff and such sources might be missed.
Effectively, this criterion implies almost a factor of two larger linear-size cutoff 
for FR-{\sc i} sources in any redshift bin. 
 
For the adopted cosmology, the selection criteria imply that our giant radio source sample
will be complete for redshifts $z \le 0.13$: all edge-brightened giant 
sources with linear size exceeding 
0.7~Mpc are detectable within the corresponding $3.5 \times 10^{7}$~Mpc$^{3}$ comoving 
volume. At redshifts exceeding 0.13, only sources with progressively larger linear size would 
be selected.  Powerful double radio sources with edge-brightened structure have radio power
exceeding about $1.3 \times 10^{25}$~W~Hz$^{-1}$ at 843~MHz, corresponding to the 
FR-{\sc i}/FR-{\sc ii} break power \citep{Ow94}.  In our adopted cosmology such 
sources would be detectable in SUMSS with flux density exceeding 0.3~Jy if they are at 
redshifts $z \le 0.13$.
 
\section{ATCA radio observations of the candidate list}

We observed the candidate list of 35 large angular-size sources at higher resolution 
using the ATCA for the purpose of detecting compact radio cores 
that could guide the optical identification process. 
The ATCA observations were also aimed at examining the compact structures in the 
individual components as a means of identifying cases where the close components were
independent sources or chance alignments.

All 35 sources were observed in 24-hr observing sessions using the 6A and 6D E-W array
configurations of the ATCA during 2001 November--December (Table~1).  The multi-channel continuum 
observations were made in the 20-cm band 
using a pair of 128-MHz bands centered at 1344 and 1440~MHz.
Every source was observed at 4 or 5 widely spaced 
hour angles, each of about 6-min duration.  The data 
were reduced in MIRIAD using an automated script with 
parameters optimized for imaging point sources. 
Multi-frequency synthesis was used to make a combined image using the data from the two adjacent 
frequencies; the images had beam FWHM in the range $5\arcsec$--$10\arcsec$.
The rms noise in the deconvolved images varied between  
0.2--1.0 mJy~beam$^{-1}$; the images were dynamic-range-limited owing to the sparse
visibility coverage.

Cores were undetected or could not be unambiguously identified in five cases
(SGRSC~J0020$-$7321, J0622$-$5938, SGRS~J0326$-$7730,  J0810$-$6800 and J1259$-$7737). 
We re-observed these five sources in the 6-cm band using the ATCA in the 6A
array configuration during 2002 November (see Table~2).  These multi-channel continuum observations
were made using a pair of 128-MHz bands centred at 4800 and 5952~MHz.
Each target was observed for a total of 4~hr with the total time
distributed over hour angle as before.  
Bandwidth synthesis was used to image the visibility data in MIRIAD; the images had beam 
FWHM in the range $2\arcsec$--$3\arcsec$.
The rms noise in the resulting images was about 0.03~mJy~beam$^{-1}$. 
Excepting SGRSC~J0622$-$5938, the observations showed compact cores between 
the diffuse radio components in all the doubtful cases. 

In sources of large angular size the region between the lobes where one 
might expect to find the radio core that is coincident with an optical 
counterpart is large and, therefore, there might be ambiguities in the 
identification of the host.  In some cases supporting evidence such as 
connected lobe emission or jet-like features might provide confirmation for 
the identification. Examining our high resolution ATCA images that were used 
for detecting radio cores of the giant radio galaxy candidates, we find that 
there is on the average about 1 compact radio source coincident with an 
optical galaxy in regions of about 140 square arcmin (avoiding the regions 
close to the giant radio sources). On the average, the sky area that we 
searched for optical counterparts of individual giant radio source candidates 
is about 5 square arcmin.  We infer that the probability of occurrence of 
a random compact object in the area where we look for IDs is 0.04 and that 
ambiguities might be expected in only $4\%$ of the cases.

In five sources (SGRSC~J0200$-$6007, J0603$-$5429, J1959$-$6402, J2150$-$6210 and
J2253$-$5813), the angular separation between compact components that were detected 
at the ends of the extended SUMSS components was less than $5\arcmin$ and, therefore,
these sources were excluded from further consideration for our giant radio source sample. In 
SGRSC~J0020$-$7321 and J0129$-$6433 the ATCA observations revealed a compact component in 
only one lobe and in SGRSC~J2228$-$5600 no hotspots were seen; the ends of these sources as seen on SUMSS 
images were $\la 5\arcmin$ and they would not be included in the complete sample.  However,
they were retained for the optical spectroscopy.

The connectedness between the close components in SGRS~J0400$-$8456,
SGRSC~J0414$-$6933 and J1920$-$7753
was not established in the SUMSS images; therefore, we made somewhat higher
resolution 20-cm band ATCA observations of these sources in 1.5-km array configurations.
The SW component of SGRSC~J1920$-$7753 was observed to have a triple structure and is
probably an independent radio source. 
The northernmost component of SGRSC~J0414$-$6933 appeared to be a separate unresolved 
source and the remainder a triple of angular size less than $5\arcmin$.
Therefore, these two sources are rejected by our selection criteria.
Two other sources (SGRSC J0622$-$5938 and J0745$-$7732) were rejected because an examination
of their SUMSS extended structure together with the ATCA compact components
suggested that these source structures are peculiar and are very unlike our expectations for
double radio sources.  The details of these observations 
are given below in Section~5 and in the Appendix as part of the notes on individual sources.

Composite images made by overlaying ATCA radio contours on SuperCOSMOS
sky survey (SSS) greyscales
were examined for any optical counterparts.
In some cases, the optical images showed galaxies coincident with 
compact radio components located within each of the close pair of extended
radio sources. These were deemed to be separate radio sources created by the
separate optical identifications and not related 
double structures created by twin beams from a single AGN.
There were 4 such candidates 
(SGRSC~J0152$-$8020, J0534$-$8203, J0551$-$5655, J2222$-$5617)
that were rejected for this reason. 

There were, in all, 22 giant radio source candidates that satisfied the criteria of angular 
size, declination and latitude.

\section{Optical spectroscopy of the host galaxies}

The redshifts of six of the 22 giant radio source candidates had been
measured previously \hfill\break
(SGRSC~J0020$-$7321, J2336$-$8151, SGRS~J0143$-$5431, J1919$-$7959, and J2159$-$7219  
by \citet{Bo00}; SGRS~J0515$-$8100 by R.~Subrahmanyan et al. (2005, in preparation)). 
The spectra in \citet{Bo00} were obtained in service mode at the 3.9-m Anglo-Australian
Telescope.  The remaining 16 were observed in multiple sessions using the Dual Beam 
Spectrograph (DBS) on the ANU 2.3-m telescope \citep{Ro88} to 
obtain optical spectra for measuring redshifts.  
A journal of the optical spectroscopy is in Tables~3 and 4.

We were initially allocated a total of 7 nights of dark time. 
These were in two separate observing
runs in 2002 August (4 nights) and 2002 November (3 nights) and 14 
objects were observed in these sessions (Table~3). Only the last night 
in the August session and the later two nights in the November session were without 
cloud for most of the night and had average seeing of $\approx 1\farcs7$. 
The other nights were mostly lost due to cloud.   

Three of the objects were too faint to give reliable redshifts.
Additionally, the identifications of two other sources were revised on the
basis of ATCA observations that were scheduled following the optical spectroscopy.
We reobserved three of these candidates---SGRS~J0810$-$6800 (new id), 
J1911$-$7048 and J1946$-$8222---along with SGRS~J1259$-$7737 that had not
been observed in the earlier sessions, on two additional nights in
2004 April.  These nights were clear with average seeing of $1\farcs5$.
The two other sources---SGRS~J0237$-$6429 and J0326$-$7730 
(new id)---were observed in a separate run in 2004 September.  
These later observations are listed in Table~4.

A low resolution grating, 158R lines/mm was used, giving a wavelength coverage of
3600--10900~\AA\ with a spectral resolution of 8~\AA\ ($\sim 2$ pixels). The red arm of the 
DBS was used for the observations listed in Table~3 whereas
the blue arm was used for the observations listed in Table~4. 
Most objects were observed in a single
2000-s exposure; however, fainter objects required longer
integration times. To calibrate for atmospheric absorption a smooth spectrum star was
also observed at roughly the same zenith angle as the target observations.
In the 2002 observing runs each target and smooth spectrum standard frame was sandwiched 
between two short exposures of a Ne-Ar arc for wavelength calibration; a Cu-Ar arc 
was used in the later runs.
A slit width of $2\arcsec$ was used for the target observations as well as 
the arc exposures.  

The standard IRAF reduction package was used for the analysis. The raw
2-dimensional spectra were corrected for CCD bias after which they were
flat-fielded using averaged dome and sky flats. In extracting the 
1-dimensional spectra of the targets the sky background was averaged over
nearby rows on either side of the target trace to minimise effects due to curvature
in the spatial axis. The cosmic rays were removed by setting the flux ratio parameter for
a given flux threshold using COSMICRAYS; 
the flux ratio versus threshold plot of the CCD frame was examined for setting 
the values. This method worked well enough up to a certain flux ratio but 
beyond that it was nearly impossible to reduce the value further without also affecting 
the target spectrum adversely.
Objects observed in the sessions listed in Table~4 had multiple exposure frames
and, after calibration, these were combined using SCOMBINE; the resulting
spectra were relatively free of artefacts.  
The 1-dimensional spectra were divided by
the normalised fit to the continuum of a smooth spectrum star 
to remove atmospheric absorption.   

We used the IRAF task RVIDLINES for determining the redshifts. 
We checked for broad consistency between the 
redshift and the apparent magnitude of the target. 
Several objects were independently analysed by two of us, using different software packages, 
and we obtained identical results: this gives confidence in the redshift
estimates.

The redshifts were used to derive the linear sizes of the 22 candidates.
Four sources
(SGRSC~J0020$-$7321, J0129$-$6433, J2228$-$5600 and J2336$-$8151) 
had linear size
$< 0.7$~Mpc and were excluded from the giant radio 
source sample on this basis.  We were unable to obtain a reliable spectrum for
SGRS~J1259$-$7737 because of the faintness of the optical host; assuming that the
faintness implies a high redshift, we have chosen to retain 
this source in our giant radio source sample (see the notes on this source in Section~5.1).

\section{Results}

Table~5 presents the final sample of 18 radio sources satisfying the 
criteria: $\left\vert b \right\vert > 12\fdg5$, $\delta < -50\degr$, 
angular size $> 5\arcmin$ and linear size larger than $0.7$~Mpc;
the table lists some observed properties of the sources.  
In order to estimate the angular sizes of the giant radio sources, we measured the
peak-to-peak separation using the higher resolution ATCA images in those cases
where hotspots were detected at both ends, and used the SUMSS 843~MHz maps for the
remaining sources.
Table~6 lists the positions and flux densities of the radio cores in these 18 giant radio sources.
The radio sources in the initial list of candidates that were
deemed not to be giant radio sources on the basis of the radio-optical follow-up
observations are listed separately in the Appendix, along with their 
observational data.

The composite radio images of all 18 giant radio sources are given in Figs.~1--18; 
contours of the higher resolution ATCA images are shown using thin lines 
and the low-resolution SUMSS images are shown using thick
contours; these are shown overlaid on grey scale representations of 
SSS optical fields. 

The optical spectra of the host galaxies of the giant radio sources are shown in Fig.~19. The
spectra are not flux calibrated. The reduced spectra often have spurious 
features as a result of errors in sky subtraction and cosmic rays and we 
have identified some of these in the panels
of Fig.~19.   In Table~7 we list the lines identified in the different objects and give the
estimated redshifts derived from fits to these lines. 

In Table~8 we list some source properties derived from the 
observations: linear sizes, core powers, 
the integrated source powers and host absolute magnitudes. 

\subsection{Notes on individual sources}

\subsubsection{SGRS~J0047$-$8307 (Fig~1)}  In the SUMSS 843~MHz image the source is seen as a 
triple; there are noticeable gaps in the radio emission between 
the lobes and the core. The lobe with the smaller separation from the core 
is also the brighter of the two lobes. The ATCA image shows two compact components in the 
NW lobe; however, in the southern lobe there is only a weak component close to the peak of
the extended emission. The extension seen in the SE lobe at 843~MHz 
towards NW might be confused with emission associated with a relatively bright, 
$b_{j} = 17.8$, elliptical galaxy that is host to a compact $S_{1.4} = 6.4$~mJy source. 
The host galaxy, which is coincident with the radio core, has a close neighbour $5 \arcsec$ 
to the south and another $12 \arcsec$ to the west. The optical spectrum has low signal-to-noise 
ratio; the identification of [OIII]5007 is uncertain because of its proximity to night sky [OI]6300.

\subsubsection{SGRS~J0143$-$5431 (Fig~2)} The SUMSS 843-MHz image shows a 
relaxed morphology for the two lobes and the higher-resolution 1.4-GHz ATCA image 
shows no compact features associated with the lobes. 
The ATCA image (Fig.~2b) shows a radio core and, interestingly, this image also shows two 
compact features on either side of the core and along the source axis: this triple structure at 
the core together with the relaxed outer lobes 
suggest the possibility that SGRS J0143-5431 might be a case where an inner 
double has been created by a restarting of the beams from the central engine.
The host galaxy has a few,
relatively faint neighbours within $15 \arcsec$ radius 
and a bright galaxy $30 \arcsec$ to the SE. 
The close faint neighbour $5 \arcsec$ to the West is a blue object and the 
galaxy $12 \arcsec$ to the North is a spiral. The optical spectrum of the host galaxy 
was obtained by \citet{Bo00}. 

\subsubsection{SGRS~J0237$-$6429 (Fig~3)}  The SUMSS image shows two lobes with
surface brightness decreasing towards the centre and prominent gaps in the 
emission between the core and the two lobes. The ATCA image shows a weak core and 
compact features at the ends of both lobes. The closer lobe is also the brighter of 
the two lobes. The host galaxy does not have any close neighbours. 
The optical spectrum is very noisy and the redshift is uncertain.

\subsubsection{SGRS~J0326$-$7730 (Fig~4)} This source has an unusual structure. 
The SUMSS image shows two extended components of which the 
SE is significantly stronger. The higher resolution 1.4-GHz ATCA image reveals a 
bright elongated structure in this SE lobe. 
Although edge-brightened, there is no compact
unresolved feature towards the outer end of this lobe. The ATCA image also shows a 
compact component at the peak of the NW lobe.
The 4.8-GHz ATCA image (Fig.~4b) shows a weak core coincident with
a faint galaxy that is located within the SUMSS contours of the stronger lobe and close
to the line joining the compact features in the two lobes.  The host galaxy has several
faint galaxies in its neighbourhood.

\subsubsection{SGRS~J0331$-$7710 (Fig~5)}  This radio source has the 
largest linear size in the sample and is extremely asymmetric in 
lobe separation as well as lobe brightness and flux density. 
Both lobes are highly elongated with high axial ratios and both elongated lobes
are collinear.  The two lobes are edge brightened; however, the contrast in surface
brightness along the lobe axes is relatively weak in this source.
In the ATCA image the two lobes are completely resolved and a weak core is detected
at the inner end of the northern lobe (Fig.~5b); there is a significant emission gap of
almost 600~kpc between the core and the southern lobe.
The host galaxy is fairly isolated except for a close 
neighbour $\sim 10 \arcsec$ to the SW. There is a prominent halo associated 
with the host galaxy that appears asymmetric in the blue image, with an extension to the north.

\subsubsection{SGRS~J0400$-$8456 (Fig~6)}  
The SUMSS image shows several emission peaks that are not all collinear. 
The ATCA image shows a central source coincident with a galaxy; this radio
core has twin jets on either side (Fig.~6b).  The SUMSS peaks to the north and 
south of the central component have compact features in the ATCA images.
In the SUMSS image there is an extension from the northern peak towards west
that is oriented nearly orthogonal to the inner radio axis.
There is a similar bend in the southern lobe towards SE. Compact structure is seen associated 
with the southernmost component in the high resolution ATCA image. 
There is no optical identification at this location. We are not certain if this 
component is a part of the giant radio galaxy and do not include it in the 
size determination. A low-resolution 1.4-GHz ATCA image made using 
the 1.5~km array is shown Fig.~6(c); the connectedness of the differrent components of 
the source is seen more clearly in this image. We have
conservatively estimated the angular size to be the sky angle between the outer components
of the source in Fig.~6(c). The source is edge-darkened,  it is also one of the 
lowest power sources in the sample and the source may be classified as FR-{\sc i} type. 
The multiple peaks on either side of the core have 
Z-symmetry: this is suggestive of multiple activity episodes in a precessing jet.
The host is a relatively bright $b_{j} = 16.9$ galaxy with
several close neighbours within $1\arcmin$ radius.

\subsubsection{SGRS~J0515$-$8100 (Fig~7)}  This double radio source has extremely low
surface brightness lobes that have an extremely low axial ratio; the higher resolution ATCA
observations have detected a faint radio core coincident with a galaxy that
is located at the inner end of the southern lobe (Fig.~7b). There is an emission gap
between the core and the northern lobe. This source appears to be a giant fat double
radio source and is being studied in detail by R.~Subrahmanyan et al. 
(2005, in preparation). 
The optical spectrum of the host galaxy was obtained by Vincent
McIntyre and Carole Jackson in 1999 using the DBS at the SSO 2.3-m
telescope.  

\subsubsection{SGRS~J0631$-$5405 (Fig~8)} This giant radio source is a quasar.
The SUMSS image shows a pair of lobes; the higher
resolution ATCA image shows a strong radio core coincident with a star-like object.  
The ATCA image shows compact structure at the end of the western lobe
whereas the eastern lobe appears resolved. 
The western lobe has a smaller extent from the core and also
has a higher surface brightness.  There is an emission gap between the core
and the lower-surface-brightness eastern lobe.  The quasar has been catalogued 
in the Deep X-ray Radio Blazar Survey \citep{Pe98} and is  
a known X-ray source in the ROSAT PSPC database of point sources. 

\subsubsection{SGRS~J0746$-$5702 (Fig~9)} In the SUMSS this radio source appears as a bright extended 
central source flanked by much fainter lobes. In the higher resolution ATCA image 
the central source is resolved into a strong core and a jet-like feature to the SW.  
Both lobes are resolved in the ATCA observations and no compact structure is observed. 
The jet appears directed 
towards the fainter of the two lobes. The radio structure at 843~MHz can be
classified as FR-{\sc i} type given the edge-darkened lobes; however, the detection
of a one-sided partial jet some distance from the core is unusual and indicates
that this source might be another example of a restarting jet within relict lobes. 
There is a neighboring galaxy situated $\sim 4 \arcsec$ from the host galaxy in a 
position angle perpendicular to the radio axis of the source.

\subsubsection{SGRS~J0810$-$6800 (Fig~10)} The SUMSS image shows two lobes; the NW lobe is 
edge-brightened and has the higher surface brightness. A compact source is detected at
the peak of the NW lobe in the ATCA 1.4-GHz image; the SE lobe is completely resolved. 
The ATCA image at 4.8~GHz (Fig.~10b) 
revealed a faint radio core coincident with a star-like object that is located 
on the line joining the outer ends of the two lobes. The lobes do not, however,
extend along this line and appear directed to the north of the axis in both cases.
The host object has very broad H$\beta$ and H$\alpha$ emission lines and is a quasar.
The host object is star-like in blue; 
however, in the red SSS images
it is surrounded by three faint patches of emission.
An unusual aspect of this giant radio source is that although it has a broad-line host, 
the radio core is weak.  An unusual feature of the spectrum is the associated blue-shifted 
absorption in the Balmer lines, indicating the presence of recent star formation.

\subsubsection{SGRS~J0843$-$7007 (Fig~11)} The radio structure of this giant radio source resembles SGRS~J0746$-$5702
in that it has a central, bright component straddled by two extended lobes with low surface brightness.
In the higher resolution ATCA image only a central radio core is detected; 
the two low-surface-brightness lobes are completely resolved. The host galaxy 
has a close neighbour $\sim 10\arcsec$ to the north.  This source might be a relic of a giant FR-{\sc ii} radio
source in which the beams from the central engine no longer energize the lobes.

\subsubsection{SGRS~J1259$-$7737 (Fig~12)} This giant radio source has prominent lobes and is clearly seen to
be an edge-brightened double radio source. The two lobes extend all the way towards 
each other without any emission gap. The 1.4-GHz ATCA high-resolution image detected compact hotspots at the
ends of both lobes and later 4.8-GHz ATCA imaging detected an additional weak compact
core close to the centre of the radio source (Fig.~12b). The core is
coincident with a faint object at the plate limit. The sky region in which the host galaxy 
lies appears relatively devoid of optical objects suggesting dust obscuration along this line of sight. 
In spite of $5 \times 2000$~s exposure on the 2.3-m telescope we 
were unable to obtain a firm redshift for this galaxy because of its faintness.  
A K-band image of the field was obtained for us by Helen Johnston with the Cryogenic Array 
Spectrometer/Imager 
(CASPIR) on the ANU 2.3-m telescope at Siding Spring Observatory. A host galaxy is 
clearly detected 
at the location of the radio core. We estimate its K magnitude to be 16.2 
$\pm0.2$; using 
the K-z relation for radio galaxies we estimate its redshift to be $>0.3$. Its linear size 
is estimated to be $>1.5$ Mpc.

\subsubsection{SGRS~J1336$-$8019 (Fig~13)} In the SUMSS this double radio
source appears to have an edge-brightened 
structure with a high axial ratio. The higher resolution ATCA image shows a chain of weak 
compact features that trace the chain of peaks observed in SUMSS indicating the presence of a weak
two-sided jet along the central line of this radio source.  The strongest among the
emission peaks is the central core component and this coincides with the southern member 
of a close pair of objects (Fig.~13b) which were partly blended on the spectrograph slit. 
The presence of H$\alpha$ absorption in the host galaxy spectrum indicates that the northern object is a star. 
The ATCA image also detects a feature at the leading edge of the southern lobe; however, there is no
recognizable equivalent feature at the leading edge of the northern lobe. 

\subsubsection{SGRS~J1728$-$7237 (Fig~14)} The source is a triple in both the SUMSS and ATCA 
images. There are large emission gaps between the lobes and the core. 
This is an asymmetric source with the northern lobe closer to the core as compared to
the southern lobe.
The ATCA image detects a radio core that is displaced $\sim 7\arcsec$ from a
bright star and coincident with a faint galaxy, which has a prominent narrow emission-line spectrum.
There are several faint galaxies visible in the neighbourhood of the host. 

\subsubsection{SGRS~J1911$-$7048 (Fig~15)} The large-scale radio structure indicates a 
low-surface-brightness edge-brightened radio source with a continuous bridge
connecting the two lobes. In the higher resolution
ATCA image only a core is detected at the centre of the source and coincident with a
faint galaxy; the rest of the lobe structure is  completely resolved. 

\subsubsection{SGRS~J1919$-$7959 (Fig~16)} This giant radio source was observed earlier 
by \citet{Su96}.  The SUMSS image shows a pair of prominent lobes;
the ATCA image detects structure in both these lobes.  Additionally, the
ATCA image from \citet{Su96} detects a weak core between the two lobes and coincident 
with a faint $b_{j}=22.7$ galaxy.  Inspection of the 
digitized plates reveals significant dust obscuration. The optical spectrum was obtained 
by RWH in 1994 using the Faint Object Red Spectrograph on the Anglo-Australian Telescope.

\subsubsection{SGRS~J1946$-$8222 (Fig~17)} The high resolution ATCA image of this source
shows a compact, weak core component coincident with a faint host galaxy; the core
is straddled by a pair of partial jet-like features that are symmetrically located
on either side of the core.  The SUMSS image shows extended emission around this 
central structure as two lobes that have a high axial ratio.  The two ends of the 
source are asymmetrically located with respect to the core.
There are no prominent compact hotspots seen at the ends of the source 
in the 1.4-GHz ATCA image.  The inner partial jets might be a case of
restarting beams in this giant radio source.
The host galaxy is barely seen in the blue, while in the
red the host appears as a point-like feature with faint extended optical
emission. We twice attempted to obtain a redshift for this galaxy but poor observing 
conditions allowed only a tentative estimate.

\subsubsection{SGRS~J2159$-$7219 (Fig~18)} In the SUMSS this source has an 
edge-brightened outer structure and a central triple structure with emission
gaps between all components.  There is a larger emission
gap between the northern lobe and the core. 
In the ATCA 1.4-GHz image a compact radio core is 
detected at the location of the central component and this coincides with a 
bright galaxy (Fig.~18b). At this higher frequency and resolution the only other structure seen is the 
stronger southern component of the central triple.  The host galaxy has a prominent 
halo; fainter companions are seen in its neighbourhood and the host is likely 
to be in a cluster environment. The optical spectrum was obtained by \citet{Bo00}.

\section{Discussion}

SUMSS has an rms noise of about 1~mJy~beam$^{-1}$ and
a 5$\sigma$ surface brightness detection limit of about 12~mJy~arcmin$^{-2}$.  With a beam
that has a FWHM close to $1\arcmin$, it is the best survey today for finding 
low redshift giant radio sources
in the southern sky.  The sample of giant radio sources that we have compiled from SUMSS
and presented in this paper is the only complete sample in the south.  

We find 18 giant radio sources in the 2100 square degrees of search area.  There are 4 giant
radio sources in this sample with redshifts $z \le 0.13$: these are in the comoving search volume of 
$3.5 \times 10^{7}$~Mpc$^{3}$ for which the survey is expected to detect all edge-brightened 
giant radio sources with projected linear size exceeding 0.7~Mpc.  This implies
that the space density of giant radio galaxies is about $10^{-7}$~Mpc$^{-3}$  at the survey 
sensitivity or that a giant radio source is expected per (215~Mpc)$^{3}$ in the local universe.

The space density of powerful FR-{\sc ii} radio sources in the local universe 
is about $10^{-7}$~Mpc$^{-3}$ \citep{Pa92}.
Of the four giant radio sources we detect at $z \le 0.13$, 
SGRS~J0400$-$8456 is a FR-{\sc i} radio galaxy while the remaining three 
(SGRS~J0515$-$8100, J0746$-$5702 \& 
J2159$-$7219) are sources with relaxed---possibly relic---lobes and have radio luminosities
significantly below the FR-{\sc i}/FR-{\sc ii} break luminosity of $1.3 \times 10^{25}$~W~Hz$^{-1}$.
None of the giant radio galaxies in our sample at $z \le 0.13$ is a classic powerful radio source:
none of them has hotspots within the lobes and none of them is above the FR-{\sc i}/FR-{\sc ii}
break in radio luminosity.  Within the comoving search volume of 
$3.5 \times 10^{7}$~Mpc$^{3}$ where the survey is expected to detect all edge-brightened 
giant radio sources with projected linear size exceeding 0.7~Mpc, all the giants we detect
have relict lobes suggesting that {\it at low redshifts giant radio sources with relict 
lobes are more numerous than active giants in which the lobes are energized by beams from 
the centre.}  The lack of powerful edge-brightened giant
radio sources in this survey volume is consistent with the expected small number of powerful 
edge-brightened radio sources given the small search region. It is likely that the giant 
radio sources we detect at $z \le 0.13$---which have edge-brightened relict lobes---are relics of 
FR-{\sc ii} giant radio sources. The characteristics of these $z \le 0.13$ sources allow us to make 
the tentative inference that FR-{\sc ii} giant radio sources may have much shorter 
lifetimes in the active phase as compared to the time for which their fading 
relict lobes remain visible above the survey sensitivity limit.

46 giant radio sources were found in the WENSS survey area of 8100 square degrees north of
declination $\delta = +28\degr$ \citep{Sc01}.  The angular size and linear-size 
cutoffs in the selection of WENSS giants was the same as the criteria we have adopted here. 
The WENSS sensitivity varied with declination: assuming that giant radio sources have a spectral 
index $\alpha = -0.7$--$-1.0$ ($S_{\nu} \propto \nu^{\alpha}$), the WENSS has a sensitivity to these extended
sources that is somewhat better than the SUMSS sensitivity in the sky north of $\delta =+74\degr$ and
comparable in most of the survey region $+28\degr < \delta < +74\degr$. In the 600 square degrees where the
surface brightness sensitivity in the WENSS is better, the detection rate in the WENSS search
is a factor 1.7 higher.  However, in the 7500 square degrees where the surface brightness
sensitivity of the WENSS is comparable to that of the SUMSS, the detection rate in the WENSS is
only 0.6 of that in our SUMSS based search.  It may be noted that \citet{La01} 
have pointed out that as many as 8 giant sources at $\delta > +60\degr$ were missed being selected
in the compilation of \citet{Sc01}. 

Barring two giant radio sources (SGRS~J0400$-$8456 and J2159$-$7219) none of the SUMSS 
giant radio sources in the sample presented here are likely to belong to clusters. 
We examined their galaxy environments in the SSS fields and we find them to be isolated,
having close companions or in a small group of galaxies. 

Six of the eighteen giant radio sources in the sample show classic FR-{\sc ii} radio morphology 
with compact hotspots at each of the lobe ends (SGRS~J0047$-$8307, J0237$-$6429, J0631$-$5405,
J1259$-$7737, J1728$-$7237 and J1919$-$7959).  SGRS~J0810$-$6800 and J1336$-$8019 are FR-{\sc ii}s 
with compact structure at the end of only one lobe;
in both these sources the ATCA images show no compact component in the other lobe.

SGRS~J0326$-$7730 and J0331$-$7710 have large scale SUMSS structures that are highly asymmetric
in lobe flux density and lobe separation from their cores.  Of these,
SGRS~J0326$-$7730 is observed to 
have structure at the end of one lobe in the ATCA image; SGRS~J0331$-$7710 is not observed to have 
any compact structure in either lobe.

Seven sources show relaxed lobes with no compact hotspots 
at either end
(SGRS~J0143$-$5431, \hfill\break
J0515$-$8100, J0746$-$5702, J0843$-$7007, J1911$-$7048, J1946$-$8222
and J2159$-$7219). All these seven sources
with relaxed lobes are among the relatively low-radio-power sources in the sample; this is 
consistent with the interpretation that these are relic sources in which beams do not currently
feed the outer relaxed lobes.  Of these, SGRS~J0143$-$5431, J0746$-$5702, J1946$-$8222 and J2159$-$7219 have 
evidence for either one or two sided knots/jets closer to the nucleus and might be examples of
giant radio sources with relict lobes and restarting beams.   

SGRS~J0400$-$8456 is the only source in the sample that is observed to have highly-bent 
tail-like lobe structures that decrease in surface brightness away from the centre.  This source appears
to have a pair of jets close to the centre and might be a giant edge-darkened radio source.
The source has among the lowest radio powers in the sample and this is consistent with the 
FR-{\sc i} type structure.

The SUMSS is a survey with high surface brightness sensitivity; nevertheless, a large
fraction (11/18) of the giant radio sources in the sample are observed to have emission gaps
between the lobes.  Continuous bridges are observed in SGRS~J0400$-$8456
J0746$-$5702, J0843$-$7007, J1259$-$7737, J1336$-$8019, 
J1911$-$7048 and J1919$-$7959.

There are two quasars in the sample. Both the quasars
(SGRS~J0631$-$5405 and J0810$-$6800) have prominent lobes and have significantly 
more compact emission detected at the end of one of their lobes. 
While J0631$-$5405 has a bright radio core consistent with 
it being a quasar, the core in J0810$-$6800 is relatively  weak. 
In both quasars the brighter lobe is observed to be closer to the core 
and is also the one containing the hotspot. 
No jets are seen in the ATCA images associated with the cores in these sources. 
Compared to the sources in the sample with classic FR-{\sc ii} structure that have
hotspots at one or both ends of the lobes, the two quasars have wider lobes with significantly
lower axial ratio.  It may be noted here that if the quasars are significantly
fore-shortened owing to projection, their linear sizes might be significantly larger than
the estimates given in Table~8 and, therefore, these quasars are potentially the sources
with the largest linear size in the sample. The giant quasars have intermediate radio 
powers compared to other giant radio sources in our sample.

The quasar fraction that we detect (two quasars in the sample of 18 giant radio sources) 
is smaller than that expected on the basis of the unification model 
for radio galaxies and quasars \citep{Ba89}.  This might be because giant radio sources with larger
linear sizes are relatively rarer and our lower limit of $5\arcmin$ to the angular size of
sources accepted into the sample might imply a larger linear-size cutoff for quasars as compared
to radio galaxies.

Seven of the host galaxies of the giant radio sources in the sample 
have detections of narrow emission lines.   
Of these SGRS~J0326$-$7730, J1728$-$7237 and J1919$-$7959
have relatively stronger emission lines whereas the emission lines in SGRS~J0047$-$8307, 
J0237$-$6429, J0515$-$8100 and J1336$-$8019 are weaker.  The giant radio sources
in the sample with relatively higher redshifts and radio powers appear to be the ones with
emission lines in their host spectra.

Asymmetry in lobe extents appears to be a characteristic of the sample.  
We define an asymmetry parameter as the ratio of the separation from the core of the 
farther lobe to the separation from the core of the closer lobe; distances to
both lobes are measured to their farthest emission peaks.
If we exclude the two quasars that might be affected by projection because their axes 
are likely to be at relatively smaller angles to the line of sight,
11 of the 16 giant radio sources have asymmetry parameter $\ga 1.25$:
nearly  $70\%$ of the sample has radio lobes that are longer on one side by $\ga 25\%$. 
Among the most asymmetric is SGRS~J0331$-$7710 in which the southern lobe extends to 
$\ga 2.5$ times the distance of the northern lobe from the core. 

Of the 11 asymmetric giant radio sources, the closer lobe is also the brighter of the
two lobes in as many as 10 cases.  This statistically significant result, which was earlier noted by 
\citet{Sa86} for a small set of giant radio sources, has also been seen 
in other samples of giant radio sources \citep{Sc00b,La01,Ma01}. 
We do not, however, find a correlation between the lobe separation
asymmetry and the lobe brightness asymmetry: although SGRS~J0331$-$7710 is the 
most asymmetric in lobe separation and has lobes that are also very asymmetric 
in brightness it is not among the most asymmetric in lobe brightness in the sample. 

We define the misalignment angle as the supplement of the angle subtended by the two outermost
hotspots at the core.  Excluding the two quasars, the bent FR-{\sc i} SGRS~J0400$-$8456
and the sources that have lobes with small
axial ratio and no compact structure (SGRS~J0143$-$5431, J0515$-$8100, J0746$-$5702 and J0843$-$7007),
the distribution of misalignment angles shows that all the remaining 11 sources are aligned to 
within $10\degr$ over the megaparsec distance; 7 of the 11 are aligned to within $5\degr$.

The median linear size of our SUMSS sample is 1.3~Mpc and the median total power
is $5\times 10^{25}$~W~Hz$^{-1}$.  The giant radio sources in the sample 
have luminosities spread over more than two orders of magnitude 
covering the range $10^{24} < P_{total}^{843} < 5\times 10^{26}$~W~Hz$^{-1}$,
straddling the FR-{\sc i}/FR-{\sc ii} break. 
The SUMSS giant radio sources exhibit increasing core power with total power: the correlation
coefficient between the logarithm of the core power at 1.4~GHz and the logarithm of the total
power at 843~MHz is only 0.5 for our sample; however, the fit is consistent with that 
found by \citet{Gi01} for B2 and 3CR radio galaxies that cover a larger range in total
luminosity but having smaller linear size.  
In spite of its relatively low total power, SGRS~J0746$-$5702, which has a prominent partial jet
towards its SW lobe, has a core power that is well above this fit to the core power versus total 
power: this is consistent with the the earlier hypothesis that SGRS~J0746$-$5702 is a restarted source. 
The only giant radio source in the sample with strong, narrow emission lines
(SGRS~J1728$-$7237) also has the most powerful radio core.  
Surprisingly, the giant radio source in our sample with the lowest core power is a
quasar (SGRS~J0810$-$6800).     

\section{Summary}

We have presented a complete sample of 18 giant radio sources compiled from the SUMSS; the sample
satisfies the selection criteria: Galactic latitude $\left\vert b \right\vert > 12\fdg5$,
declination $\delta < -50\degr$ and angular size $ > 5\arcmin$. All the 
sources have projected linear size larger than 0.7~Mpc (assuming $H_{o}$ = 
71~km~sec$^{-1}$~Mpc$^{-1}$). Higher resolution ATCA radio observations at 1.4~GHz have been 
used to identify radio cores and hotspots and distinguish any unrelated radio sources in the original
candidate list. This facilitated optical identification and subsequent optical spectroscopy using 
the DBS at the 2.3~m SSO telescope and resulted in near complete redshift 
information for our sample. The redshift range of the sample is 0.09$-$0.48. 
The 18 giant radio sources have powers straddling the FR-{\sc i}/FR-{\sc ii} break power. 

The SUMSS giants have radio morphologies that are mostly edge-brightened although several have
lobes with no compact features in them. Asymmetric lobe separation from their cores is widely seen 
and a significant correlation exists between lobe separation and lobe brightness: the closer lobe 
is nearly always the brighter. 

We detect all giants with linear size larger than 0.7~Mpc in the nearby universe
($z \le 0.13$). However, given the selection criteria, there is inherently 
a redshift-dependent linear-size bias 
at higher redshifts. Within this survey area ($z \le 0.13$), not one of the four giants we 
detect has edge-brightened structure with compact hotspots.
All four giants have among the lowest powers in the sample and three have 
lobes with relaxed morphologies. If the three with relaxed morphologies 
are relicts of FR-{\sc ii} giants this suggests that in the nearby universe there 
could be an overabundance of giant radio sources with relict lobes as 
compared to giants whose lobe ends are currently energized by beams.

The SUMSS sample has several interesting sources. The sample contains 
one of the most asymmetric giant radio sources SGRS~J0331$-$7710 where not only is one lobe about
2.5 times closer to the core than the opposite lobe, it is also more than 2.5 times
brighter. The four giant radio sources: SGRS~J0143$-$5431, J0746$-$5702, J1946$-$8222, J2159$-$7219 
as well as SGRSC~0414$-$6933 are candidates for re-started activity:  
all have relaxed outer lobes and either 
a pair of inner lobes that might be a manifestation of new beams from the central engine or partial jets 
close to the centre.

\acknowledgments
The MOST is operated by the University of Sydney and supported in part by grants from 
the Australian Research Council.
The Australia Telescope Compact Array is part of the Australia Telescope which is funded by the
Commonwealth of Australia for operation as a National Facility managed by CSIRO.
We acknowledge the use of SuperCOSMOS, an advanced photographic plate digitizing machine at the Royal 
Observatory of Edinburgh, in the use of digitized images for the radio-optical overlays.
We thank Anton Koekemoer for discussions in the early part of
the project and for advice on some IRAF tasks. We thank Helen Johnston for the K-band image of
SGRS~J1259$-$7737.

\clearpage
\appendix

\section{CANDIDATES NOT IN THE COMPLETE
SAMPLE OF GIANT RADIO SOURCES}

In Figs.~20--28 we present 843~MHz SUMSS and higher resolution ATCA images of a selection of 
giant radio source candidates from the original list that did not satisfy one or 
both criteria: angular size $ > 5\arcmin$ and projected linear size larger than 0.7~Mpc. 
We obtained low-resolution optical spectra using the DBS on the 2.3-m SSO telescope for
the host galaxies of some of these candidates and \citet{Bo00} obtained AAT spectra for
some hosts; the spectra are presented in Fig.~29.
Table~A1 lists the prominent spectral lines and the derived redshifts. 
In Table~A2 we give the sizes and structural type for some of these radio sources. 
In Table~A3 we list their radio core positions and the 1.4-GHz core flux densities. 
For all the sources from the original giant radio source candidate list 
that were rejected from the final sample we give below brief notes. 

\subsection{SGRSC~J0020$-$7321 (Fig~20)} The SUMSS image shows a filled radio structure
over its $4\farcm8$ extent. In the higher resolution 1.4-GHz ATCA image the eastern 
lobe is completely resolved (Fig.~20a). There is structure seen in the western
lobe; however, no compact hotspot is observed. The 4.8-GHz ATCA image detects
a core at the position of a bright galaxy (Fig.~20b). Twin-jet like structures are seen 
at the core and aligned with the
source axis; the western jet appears to curve towards the bright elongated structure
in the western lobe. The host has a close 
companion $4 \arcsec$ to the NW and there is a prominent halo surrounding 
the galaxy pair. This candidate is rejected because it 
fails the angular-size criterion. The optical
spectrum was obtained by \citet{Bo00}.

\subsection{SGRSC~J0129$-$6433 (Fig~21)} The SUMSS image shows a filled radio 
structure with a straight edge along its SW boundary.  The radio structure is reminiscent of the 
structure seen in 3C430 and 3C111 where one edge of the radio galaxy has a 
straight edge and lobe emission on the opposite side extends away 
perpendicular to the radio axis. The ATCA image shows a compact 
feature that appears to be a multiple hotspot structure at the end of the SE lobe; 
the northern lobe is resolved.  There are two presumably unrelated sources seen in 
the ATCA image within the southern lobe.  The host galaxy of this source has strong narrow 
emission lines and is fairly isolated. The closest neighbours are about an arcminute 
away to the SE and SW.  The candidate fails the angular-size criterion 
and is excluded from our sample. 

\subsection{SGRSC~J0152$-$8020} In the SUMSS image (not shown) the source is seen as two extended 
components separated
by about $5\farcm5$. Both components of this candidate are detected in
the higher resolution ATCA observations in which the SW source is identified with
a faint optical object and the NE source has a double structure in position angle
$-20\degr$. The candidate is rejected because the western component is
identified and both SUMSS components are likely to be separate sources.

\subsection{SGRSC~J0200$-$6007 (Fig~22)} Two prominent lobes are observed in the SUMSS 
image. The ATCA observations detect structure in both lobes; additionally, 
a compact core is detected between the two components. 
The core is identified with a faint $b_{j}=21.2$ galaxy; we have
not obtained an optical spectrum for the host galaxy. This
candidate is rejected because its angular extent is less than $5\arcmin$. 

\subsection{SGRSC~J0414$-$6933 (Fig~23)} Four extended components are seen in the
SUMSS image of this candidate.  Higher resolution ATCA image shows a compact
source coincident with the Northern-most component.  The other three SUMSS
components are observed to constitute a separate triple radio source in the
low-resolution ATCA image (Fig.~23b): the central component of this triple 
is resolved into a $45\arcsec$
triple in the higher-resolution ATCA image (Fig.~23a) and the core is identified with 
a $b_{j}=19.5$ galaxy. The lack of compact features in the two lobes and the central
triple structure suggest restarting activity in this radio galaxy.
The optical host has a halo that is more prominent in
the red and has a sharp edge towards the SW. It has no immediate neighbours 
although it might be the member of a loose group.
The $4\farcm5$ triple radio source is a giant radio source; however,
J0414$-$6933 is not a part of the sample because its angular size is less than
$5\arcmin$. The optical spectrum for this source (Fig.~29) was obtained using the Dual Beam 
Spectrograph at the 2.3-m telescope at Siding Spring Observatory. 
The spectrum was obtained from a single 2000-s exposure on 2002 Nov 04.

\subsection{SGRSC~J0534$-$8203} This source is observed to be two separate
components in the SUMSS image (not shown); the candidate has a total angular size less than 
$5 \arcmin$ and therefore fails the selection criterion.  In the ATCA image (not shown)
the eastern component is resolved into a $1\arcmin$ triple;
there is a faint blue galaxy very close to the central
component of this triple, which is likely to be the host galaxy for the eastern SUMSS
component. The western component is 
resolved into three compact sources none of which has an optical counterpart in SSS. 

\subsection{SGRSC~J0551$-$5655} This source consists of three extended components
in the SUMSS image (not shown).  Higher resolution ATCA observations show the western component 
to be a double 
centered on an optical galaxy; additionally, compact components without optical
counterparts in SSS images are detected coincident with the other two SUMSS components. 
The candidate has an angular size less than $5 \arcmin$ and hence fails the selection criteria. 

\subsection{SGRSC~J0603$-$5429} In the SUMSS image (not shown) the source is seen as
three peaks that are connected by extended emission. In the higher resolution 
1.4-GHz ATCA observations the SE peak appears as an unresolved component without
any optical counterpart in SSS images.
The remaining two peaks are also detected: they are weaker and appear
to be coincident with two faint galaxies.
This source is rejected as a candidate because its
peak-to-peak angular size is less than $5\arcmin$. 

\subsection{SGRSC~J0622$-$5938} In the SUMSS image (not shown) the source appears as two components that
are extended roughly towards each other with a faint bridge of emission connecting them.
The ATCA detects weak components at the two peaks and these have no optical counterparts
in SSS images.  The candidate has an angular size less than 
$5 \arcmin$ and fails the selection criteria. 
 
\subsection{SGRSC~J0745$-$7732} In both the SUMSS and the ATCA images (not shown) this source
is observed to be a chain of nine unresolved radio sources arrayed along
a position angle of about $45\degr$. Some of the 
components are extended along this position angle in the SUMSS; however, 
there is no diffuse emission connecting these individual sources. Only two 
components have faint objects associated with them in the blue SSS image. 
Apart from being in a line, there is no evidence
from either the SUMSS or ATCA observations that the remaining components form a 
single source and, therefore, we do not accept this into our giant radio source sample.

\subsection{SGRSC~J1920$-$7753 (Fig~24)}  The SUMSS image shows a 
very elongated radio structure directed towards a component to the NE (Fig.~24a). 
Low-resolution 20-cm ATCA imaging revealed that the elongated
source is a triple (Fig.~24b). In 
the higher resolution ATCA image the elongated source is resolved into a series of peaks and 
the brightest (with flux density 19.4~mJy~beam$^{-1}$) is identified with a very faint object 
that is seen only in the red SSS image.  The NE component is detected as an unresolved source
without any optical ID in SSS.    Because the SUMSS components are deemed to be separate 
sources that are $< 5\arcmin$ in size, the source is excluded from the sample.

\subsection{SGRSC~J1959$-$6402 (Fig~25)}  The SUMSS image shows this source to be two edge-brightened lobes;
ATCA images show compact hotspots at the leading edges. A weak core is 
detected with a flux density of 4.3~mJy~beam$^{-1}$ at 1.4~GHz. The angular size as measured 
between the hotspots detected in the ATCA image is just under $5\arcmin$ and
the source fails the selection criteria for our sample.

\subsection{SGRSC~J2150$-$6210} In the SUMSS image (not shown) the source has a double structure, 
with significantly different flux densities, and a connecting bridge.  
The ATCA detects a compact source at the location of
the stronger NE component and a weaker source at the peak of the SW SUMSS component.
The NE source is identified with a bright elliptical galaxy and is likely to be a
separate source. The candidate is rejected as it fails the angular-size criterion.  

\subsection{SGRSC~J2222$-$5617} Two separate components appear in the SUMSS image (not shown)
and the southern component has a weak extension towards the other component. 
The ATCA observations reveal the two 
components to be separately a double and a triple. The southern triple is 
identified in SSS with an elliptical galaxy. The two SUMSS components are likely to
be two unrelated radio sources. 

\subsection{SGRSC~J2228$-$5600 (Fig~26)} The SUMSS image shows this source to be an edge-brightened 
radio source with a relatively strong core. The 1.4-GHz ATCA image shows only the core 
component as a strong unresolved source coincident with a 
relatively bright $b_{j} = 17.7$ elliptical 
galaxy. The host is likely to be in a cluster. The galaxy optical spectrum
shows the presence of moderately strong emission lines. This source has a linear size
of 428~kpc and is not a giant radio source.

\subsection{SGRSC~J2253$-$5813 (Fig~27)} The SUMSS image shows an edge-brightened
double radio source with a continuous bridge.  Higher resolution ATCA observations
detect hotspot components at both ends as well as a radio core that is 
identified with a $b_{j}=18.5$ elliptical galaxy. We have not obtained an optical
spectrum for this galaxy.  This source is not in the sample because its 
angular size $< 5\arcmin$. 

\subsection{SGRSC~J2336$-$8151 (Fig~28)} The SUMSS image shows a chain of connected 
components.  Our higher resolution ATCA image shows emission features 
associated with all except the diffuse component at the SE end of the source.
A radio core, coincident with an optical counterpart, 
is detected in an inner component and there is an indication for twin-jet 
structure close to the core. The radio structure is very asymmetric.
This source fails the linear-size criterion and hence is
not part of our giant radio source sample. The optical spectrum was obtained by \citet{Bo00}.

\clearpage

\begin{deluxetable}{lccc}
\tablewidth{0pt}
\tablecaption{Journal of ATCA observations-1.4~GHz}
\startdata
\tableline
\tableline
Name &Array& Date & Time (min) \\
\tableline
SGRSC~J0020$-$7321&  6D  &  2001~Nov~26  & 28\\
                  &  6A  &  2001~Dec~14  & 24\\
SGRS~J0047$-$8307 &  6D  &  2001~Nov~26  & 28\\
                  &  6A  &  2001~Dec~14  & 24\\
SGRSC~J0129$-$6433&  6D  &  2001~Nov~25  & 24\\
                  &  6A  &  2001~Dec~14  & 24\\
SGRS~J0143$-$5431 &  6D  &  2001~Nov~26  & 21\\ 
                  &  6A  &  2001~Dec~14  & 24\\
SGRSC~J0152$-$8020&  6D  &  2001~Nov~25  & 28\\
                  &  6A  &  2001~Dec~14  & 24\\
SGRSC~J0200$-$6007&  6D  &  2001~Nov~25  & 33\\
                  &  6A  &  2001~Dec~14  & 24\\
SGRS~J0237$-$6429 &  6D  &  2001~Nov~25  & 28\\
                  &  6A  &  2001~Dec~14  & 24\\
SGRS~J0326$-$7730 &  6D  &  2001~Nov~25  & 35\\
                  &  6A  &  2001~Dec~14  & 24\\
SGRS~J0331$-$7710 &  6D  &  2001~Nov~25  & 33\\
                  &  6A  &  2001~Dec~14  & 24\\
SGRS~J0400$-$8456 &  6D  &  2001~Nov~25  & 28\\ 
                  &  6A  &  2001~Dec~14  & 24\\
                  &  1.5A&  2004~Mar~21  & 375\\ 
                  &  1.5A&  2004~Mar~30  & 310\\  
SGRSC~J0414$-$6933&  6D  &  2001~Nov~25  & 35\\
                  &  6A  &  2001~Dec~14  & 24\\
                  &  1.5A&  2004~Mar~30  & 260\\ 
SGRS~J0515$-$8100 &  6D  &  2001~Nov~25  & 35\\ 
                  &  6A  &  2001~Dec~14  & 24\\
SGRSC~J0534$-$8203&  6D  &  2001~Nov~25  & 35\\
                  &  6A  &  2001~Dec~14  & 24\\
SGRSC~J0551$-$5655&  6D  &  2001~Nov~25  & 35\\
                  &  6A  &  2001~Dec~14  & 24\\
SGRSC~J0603$-$5429&  6D  &  2001~Nov~25  & 28\\
                  &  6A  &  2001~Dec~14  & 24\\
SGRSC~J0622$-$5938&  6D  &  2001~Nov~25  & 28\\
                  &  6A  &  2001~Dec~14  & 24\\
SGRS~J0631$-$5405 &  6D  &  2001~Nov~25  & 28\\
                  &  6A  &  2001~Dec~14  & 24\\
SGRSC~J0745$-$7732&  6D  &  2001~Nov~25  & 38\\
                  &  6A  &  2001~Dec~14  & 36\\
SGRS~J0746$-$5702 &  6D  &  2001~Nov~25  & 35\\
                  &  6A  &  2001~Dec~14  & 36\\ 
SGRS~J0810$-$6800 &  6D  &  2001~Nov~25  & 35\\
                  &  6A  &  2001~Dec~14  & 30\\
SGRS~J0843$-$7007 &  6D  &  2001~Nov~25  & 35\\
                  &  6A  &  2001~Dec~14  & 30\\
SGRS~J1259$-$7737 &  6D  &  2001~Nov~25  & 35\\
                  &  6A  &  2001~Dec~14  & 36\\
SGRS~J1336$-$8019 &  6D  &  2001~Nov~25  & 35\\
                  &  6A  &  2001~Dec~14  & 36\\
SGRS~J1728$-$7237 &  6D  &  2001~Nov~25  & 42\\
                  &  6A  &  2001~Dec~14  & 36\\
SGRS~J1911$-$7048 &  6D  &  2001~Nov~25  & 35\\
                  &  6A  &  2001~Dec~14  & 30\\
SGRS~J1919$-$7959 &  6D  &  2001~Nov~26  & 28\\
                  &  6A  &  2001~Dec~14  & 24\\
SGRSC~J1920$-$7753&  6D  &  2001~Nov~26  & 21\\
                  &  6A  &  2001~Dec~14  & 24\\
                  &  1.5A&  2004~Apr~04  & 312\\ 
SGRS~J1946$-$8222 &  6D  &  2001~Nov~26  & 21\\
                  &  6A  &  2001~Dec~14  & 24\\
SGRSC~J1959$-$6402&  6D  &  2001~Nov~26  & 21\\
                  &  6A  &  2001~Dec~14  & 25\\
SGRSC~J2150$-$6210&  6D  &  2001~Nov~26  & 28\\
                  &  6A  &  2001~Dec~14  & 24\\
SGRS~J2159$-$7219 &  6D  &  2001~Nov~26  & 28\\
                  &  6A  &  2001~Dec~14  & 24\\
SGRSC~J2222$-$5617&  6D  &  2001~Nov~26  & 21\\
                  &  6A  &  2001~Dec~14  & 24\\
SGRSC~J2228$-$5600&  6D  &  2001~Nov~26  & 21\\
                  &  6A  &  2001~Dec~14  & 24\\
SGRSC~J2253$-$5813&  6D  &  2001~Nov~26  & 21\\
                  &  6A  &  2001~Dec~14  & 24\\
SGRSC~J2336$-$8151&  6D  &  2001~Nov~26  & 28\\
                  &  6A  &  2001~Dec~14  & 24\\

\enddata
\end{deluxetable}

\begin{deluxetable}{lccc}
\tablewidth{0pt}
\tablecaption{Journal of ATCA observations-4.8~GHz}
\startdata
\tableline
\tableline
Name &Array& Date & Time (min) \\
\tableline
SGRSC~J0020$-$7321&  6A &  2002~Nov~22  & 227\\ 
SGRS~J0326$-$7730 &  6A &  2002~Nov~22  & 268\\
SGRSC~J0622$-$5938&  6A &  2002~Nov~22  & 152\\
SGRS~J0810$-$6800 &  6A &  2002~Nov~22  & 160\\ 
                  &  6C &  2004~May~15  & 160\\ 
SGRS~J1259$-$7737 &  6A &  2002~Nov~22  & 170\\ 

\enddata
\end{deluxetable}

\begin{deluxetable}{lccc}
\tablewidth{0pt}
\tablecaption{Journal of optical spectroscopy observations-I}
\startdata
\tableline
\tableline
Name & Magnitude $(b_{j})$ & Date & Exposure (s) \\
\tableline
SGRS~J0047$-$8307 &    20.2 &     2002 Aug 04       &     2000       \\
                 &         &     2002 Nov 05       &     2000       \\
SGRSC~J0129$-$6433 &    18.5 &     2002 Aug 02       &     2000       \\
SGRS~J0237$-$6429 &    20.9 &     2002 Aug 04       &     2000       \\
                 &         &     2002 Nov 03      &     2000       \\
SGRS~J0331$-$7710 &    18.2   &     2002 Aug 04      &     2000       \\
                 &         &     2002 Nov 04      &     2000       \\
SGRS~J0400$-$8456 &    16.9   &     2002 Nov 04       &     2000       \\
                 &         &     2002 Nov 05      &     2000       \\
SGRS~J0631$-$5405 &    16.9 &     2002 Nov 04      &     2000       \\
SGRS~J0746$-$5702 &    18   &     2002 Nov 04       &     2000       \\
SGRS~J0810$-$6800 &    --   &     2002 Nov 05       &     2000       \\
SGRS~J0843$-$7007 &    17.7 &     2002 Nov 05       &     2000       \\
SGRS~J1336$-$8019 &    16.7 &     2002 Aug 03       &     2000       \\
SGRS~J1728$-$7237 &    21   &     2002 Aug 01       &     2000       \\
                 &         &     2002 Aug 02      &     2000       \\
SGRS~J1911$-$7048 &    19.4 &     2002 Nov 05      &     2000       \\
SGRS~J1946$-$8222 &    22.3 &     2002 Aug 04       &     2000       \\
SGRSC~J2228$-$5600 &    17.5 &     2002 Aug 02      &     2000       \\
\enddata
\end{deluxetable}

\begin{deluxetable}{lccc}
\tablewidth{0pt}
\tablecaption{Journal of optical spectroscopy observations-{\sc ii}}
\startdata
\tableline
\tableline
Name & Magnitude $(b_{j})$ & Date & Exposure (s) \\
\tableline
SGRS~J0237$-$6429&   20.9  &     2004 Sep 13 &     $2\times 1500$ \\        
SGRS~J0326$-$7730&   20.7  &     2004 Sep 13 &     $2\times 1500$ \\           
SGRS~J0810$-$6800&   16.3  &     2004 Apr 15 &     $4\times 2000$ \\   
SGRS~J1259$-$7737&   22.4  &     2004 Apr 15 &     $5\times 2000$ \\
SGRS~J1911$-$7048&   19.4  &     2004 Apr 14 &     $3\times 2000$  \\
SGRS~J1946$-$8222&   22.3  &     2004 Apr 14 &     $4\times 2000$  \\
\enddata
\end{deluxetable}

\clearpage

\begin{deluxetable}{lcccc}
\tablewidth{0pt}
\tablecaption{Observed properties of the sample of 
18 SUMSS giant radio sources}
\startdata
\tableline
\tableline
 Name          & Redshift & Angular size & Radio structure &  
Total flux density \\
                    &     & (arcmin)     & FR-{\sc i}/FR-{\sc ii}      & 
$S_{tot}^{\rm 843~MHz}$ (mJy) \\
\tableline
SGRS~J0047$-$8307& 0.2591    &      5.6           &  FR-{\sc ii}          & 543   \\
SGRS~J0143$-$5431& 0.1791   &      5.5           &  FR-{\sc ii}           & 217    \\
SGRS~J0237$-$6429& 0.364:    &      6.6           &   FR-{\sc ii}         & 145   \\
SGRS~J0326$-$7730& 0.2771   &      5.3           &   FR-{\sc ii}          & 357 \\
SGRS~J0331$-$7710& 0.1456    &     17.7           &   FR-{\sc ii}         & 687     \\
SGRS~J0400$-$8456& 0.1037   &      6.3           &  FR-{\sc i}            & 175   \\
SGRS~J0515$-$8100& 0.1050    &      7.5           &   FR-{\sc i}/FR-{\sc ii}    & 264     \\
SGRS~J0631$-$5405& 0.2036   &      5.2           &   FR-{\sc ii}         & 694   \\
SGRS~J0746$-$5702& 0.1300   &      5.5           &   FR-{\sc i}          & 194    \\
SGRS~J0810$-$6800& 0.2311   &      6.5           &   FR-{\sc ii}         & 271    \\
SGRS~J0843$-$7007& 0.1381   &      5.6           &   FR-{\sc i}          & 203   \\
SGRS~J1259$-$7737&  $>0.3$\tablenotemark{a}    &      5.7          &   FR-{\sc ii}         & 439    \\
SGRS~J1336$-$8019& 0.2478   &     10.1           &   FR-{\sc ii}         & 483  \\
SGRS~J1728$-$7237& 0.4735   &      6.2           &   FR-{\sc ii}         & 214  \\
SGRS~J1911$-$7048& 0.2152   &      6.0           &   FR-{\sc ii}         & 238    \\
SGRS~J1919$-$7959& 0.3462    &      6.0           &   FR-{\sc ii}         & 1262   \\
SGRS~J1946$-$8222& 0.333:     &      7.4           &   FR-{\sc ii}         & 219    \\
SGRS~J2159$-$7219& 0.0974   &      9.2           &   FR-{\sc ii}         & 142     \\
\enddata
\tablenotetext{a}{Redshift lower limit is using K=16.2 $\pm 0.2$ mag. See Section 5.1.12.} 
\end{deluxetable}

\clearpage

\begin{deluxetable}{clcccc}
\tabletypesize{\small}
\rotate
\tablewidth{0pt}
\tablecaption{Results of ATCA observations of the 
giant radio source sample}
\startdata
\tableline
\tableline
 Name           &Other names    & Associated & Core position & 
\multicolumn{2}{c}{Core flux density (mJy)} \\ 
                &              & 
SUMSS sources  &RA \& DEC (J2000 epoch) & 1.4~GHz & 4.8~GHz \\
\tableline 
SGRS~J0047$-$8307&PKS B0046$-$833 &J004756$-$830816& 00 47 55.5 $-$83 08 12.5 & 8.3   &     \\
                &PMN J0047$-$8307 &J004704$-$830722&                        &       &     \\
                &MRC 0046$-$833   &J004931$-$830914&                        &       &     \\
SGRS~J0143$-$5431&              &J014344$-$543140& 01 43 43.2 $-$54 31 39.0 & 1.7   &     \\
                &              &J014353$-$543249&                        &       &     \\
                &              &J014335$-$542919&                        &       &     \\
SGRS~J0237$-$6429&PMN J0237$-$6428&J023655$-$642751& 02 37 09.9 $-$64 30 02.2 & 2.0   &     \\
                &              &J023728$-$643246&                        &       &     \\
SGRS~J0326$-$7730&PMN J0326$-$7730&J032514$-$772928& 03 26 01.3 $-$77 30 16.9 & --    & 0.46\\
                &              &J032506$-$772809&                        &       &     \\
                &              &J032618$-$773039&                        &       &     \\
SGRS~J0331$-$7710&PKS B0333$-$773 &J033133$-$771133& 03 31 39.8 $-$77 13 19.3 & 1.3   &     \\
                &PMN J0331$-$7709 &J033135$-$770900&                        &       &     \\
                &PMN J0331$-$7719 &J033138$-$772218&                        &       &     \\
SGRS~J0400$-$8456&PMN J0401$-$8457&J040116$-$845642& 04 01 18.4 $-$84 56 35.9 & 2.9   &     \\
                &              &J040058$-$845449&                        &       &     \\
                &              &J040058$-$845829&                        &       &     \\
                &              &J035945$-$845530&                        &       &     \\
                &              &J040145$-$850015&                        &       &     \\
SGRS~J0515$-$8100&PMN J0515$-$8101&J051541$-$810139& 05 15 55.0 $-$80 59 44.4 & 3.06  &     \\
                &              &J051611$-$810219&                        &       &     \\
                &              &J051653$-$810128&                        &       &     \\
                &              &J051548$-$810314&                        &       &     \\
SGRS~J0631$-$5405&PMN J0631$-$5405&J063159$-$540506& 06 32 01.0 $-$54 04 58.7 & 37.7  &     \\
                &MRC 0630$-$540&J063214$-$540405&                        &       &     \\
                &1WGA J0631.9$-$5404&J063222$-$540320&                        &       &     \\
                &              &J063151$-$540540&                        &       &     \\
SGRS~J0746$-$5702&PMN J0746$-$5703&J074615$-$570312& 07 46 18.6 $-$57 02 59.1 & 36.4  &     \\
                &PMN J0746$-$5702&J074631$-$570152&                        &       &     \\
                &PMNM 074521.2$-$565530&J074602$-$570434&                  &       &     \\
SGRS~J0810$-$6800&PMN J0810$-$6759&J081038$-$675911& 08 10 55.1 $-$68 00 07.7 & --    & 0.34\\
                &              &J081121$-$680109&                        &       &     \\
                &              &J081029$-$675855&                        &       &     \\
                &              &J081131$-$680200&                        &       &     \\
SGRS~J0843$-$7007&PMN J0843$-$7006&J084243$-$700653& 08 43 05.9 $-$70 06 55.6 & 12.0  &     \\
                &              &J084305$-$700655&                        &       &     \\
                &              &J084331$-$700739&                        &       &     \\
                &              &J084349$-$700729&                        &       &     \\
SGRS~J1259$-$7737&PKS B1255$-$773&J125846$-$773924& 12 59 09.0 $-$77 37 28.4 & --    & 0.33\\
                &PMN J1259$-$7736&J125931$-$773500&                        &       &     \\
                &              &J125920$-$773623&                        &       &     \\
SGRS~J1336$-$8019&PMN J1335$-$8016&J133600$-$801807& 13 35 59.7 $-$80 18 05.1 & 8.6   &     \\
                &PMN J1336$-$8022&J133601$-$801637&                        &       &     \\
                &              &J133600$-$801428&                        &       &     \\
                &              &J133624$-$802250&                        &       &     \\
SGRS~J1728$-$7237&PMN J1728$-$7237&J172828$-$723732& 17 28 28.1 $-$72 37 34.9 & 11.9  &     \\
                &              &J172806$-$723549&                        &       &     \\
                &              &J172859$-$724011&                        &       &     \\
SGRS~J1911$-$7048&              &J191052$-$704906& 19 10 55.2 $-$70 49 00.9 & 1.2   &     \\
                &              &J191120$-$704716&                        &       &     \\
                &              &J191025$-$705110&                        &       &     \\
SGRS~J1919$-$7959&PKS B1910$-$800&J191903$-$795741& 19 19 16.9 $-$79 59 34.9 & 2.6   &     \\
                &PMN J1918$-$7957&J191932$-$800145&                        &       &     \\
                &MRC 1910$-$800&J191856$-$795651&                        &       &     \\
SGRS~J1946$-$8222&PMN J1946$-$8221&J194640$-$822234& 19 46 50.5 $-$82 22 53.8 & 2.1   &     \\
                &              &J194552$-$822037&                        &       &     \\
                &              &J194818$-$822539&                        &       &     \\
SGRS~J2159$-$7219&PMN J2159$-$7220&J215908$-$721904& 21 59 10.1 $-$72 19 06.6 & 9.6   &     \\
                &              &J215918$-$722013&                        &       &     \\
                &              &J215947$-$722203&                        &       &     \\
                &              &J215827$-$721528&                        &       &     \\ 
\enddata
\end{deluxetable}

\begin{deluxetable}{lcl}
\tablewidth{0pt}
\tablecaption{Redshifts and spectral lines}
\tabletypesize{\scriptsize}
\startdata
\tableline
\tableline
Name          & Redshift   & Prominent lines  \\        
\tableline
SGRS~J0047$-$8307    & 0.2591$\pm0.0002$     & Absorption lines: FeI4384, H$\beta$, Mgb\\ 
SGRS~J0143$-$5431    & 0.1791$\pm0.0001$     & Weak emission line: [OII]3727\\
                     &            & Absorption lines: K, H, G, H$\beta$, Mgb\\ 
SGRS~J0237$-$6429    & 0.364:     & Weak emission lines: [O{\sc iii}]4959,5007\\
                    &            & Possible absorption line: H$\beta$\\
SGRS~J0326$-$7730    & 0.2771$\pm0.0002$     & Moderate strength emission lines: [O{\sc iii}]4959,5007\\
                    &            & Absorption lines: K+H\\
SGRS~J0331$-$7710    & 0.1456$\pm0.0001$     & Absorption lines: Mgb and Na\\
SGRS~J0400$-$8456    & 0.1037$\pm0.0003$     & Absorption lines: Mgb, Ca+Fe, and Na\\
SGRS~J0515$-$8100    & 0.1050$\pm0.0002$     & Weak emission lines: H$\alpha$, [N{\sc ii}]6583 and [S{\sc ii}]6716,6731\\
                    &            & Absorption lines: Mgb, FeI and Na\\
SGRS~J0631$-$5405    & 0.2036$\pm0.0001$     & Quasar. Broad emission lines: H$\beta$, H$\alpha$\\
                    &            & Narrow emission lines: [O{\sc iii}]4959,5007\\
SGRS~J0746$-$5702    & 0.1300$\pm0.0005$     & Absorption lines: Mgb and Na\\
SGRS~J0810$-$6800    & 0.2311$\pm0.0002$     & Quasar. Broad emission lines: H$\gamma$, H$\beta$ and H$\alpha$\\
                    &                        & Narrow emission lines: [O{\sc iii}]4363, [O{\sc iii}]4959,5007\\
SGRS~J0843$-$7007    & 0.1381$\pm0.0003$     & Absorption lines: Mgb and Na \\
SGRS~J1336$-$8019    & 0.2478$\pm0.0001$     & Weak emission lines: [O{\sc iii}]4959,5007\\  
SGRS~J1728$-$7237     & 0.4735$\pm0.0002$     & Strong narrow emission lines: [O{\sc iii}]4959,5007 \\
                    &            & Weak emission lines: [O{\sc ii}]3727, H$\beta$\\
                    &            & Absorption line: G\\   
SGRS~J1911$-$7048    & 0.2152$\pm0.0001$     & Absorption lines:H$\delta$, G, Mgb, Ca+Fe and Na \\
SGRS~J1919$-$7959    & 0.3462$\pm0.0002$     & Narrow emission lines: [O{\sc iii}]4959,5007 and 
possible [S{\sc ii}]6716,6731 \\
SGRS~J1946$-$8222    & 0.333:     & Absorption lines: G and Mgb \\
SGRS~J2159$-$7219    & 0.0974$\pm0.0001$     & Absorption lines: K, H, G, Mgb and Na\\
\enddata
\end{deluxetable}

\clearpage

\begin{deluxetable}{llcccc}
\tablewidth{0pt}
\tablecaption{Derived properties of the SUMSS giant radio sources}
\startdata
\tableline
\tableline
Name  & Redshift & Linear size  & 1.4-GHz core power & 
843-MHz total power & $M_{b}$ \\
   & &  (Mpc) & ($10^{24}$~W~Hz$^{-1}$) & ($10^{26}$~W~Hz$^{-1}$) & \\
\tableline
SGRS~J0047$-$8307& 0.2591    & 1.34 & 1.7   & 1.1 & $-$20.3 \\
SGRS~J0143$-$5431& 0.1791   & 0.99 & 0.15  & 0.19 & $-$20.6 \\
SGRS~J0237$-$6429& 0.364:    & 2.24 & 0.9   & 0.64 & $-$20.5 \\
SGRS~J0326$-$7730& 0.2771   & 1.68 & 0.11\tablenotemark{a}  & 0.84 & $-$20.0 \\
SGRS~J0331$-$7710& 0.1456    & 2.67 & 0.07  & 0.38 & $-$21.0 \\
SGRS~J0400$-$8456& 0.1037   & 0.71 & 0.08  & 0.05 & $-$21.5 \\
SGRS~J0515$-$8100& 0.1050    & 0.86 & 0.08  & 0.07 & $-$21.1 \\    
SGRS~J0631$-$5405& 0.2036   & 1.03 & 4.4   & 0.81 & $-$23.1 \\
SGRS~J0746$-$5702& 0.1300   & 0.76 & 1.6   & 0.08 & $-$20.9 \\
SGRS~J0810$-$6800& 0.2311   & 1.28 & 0.05\tablenotemark{a}  & 0.42 & $-$23.9 \\
SGRS~J0843$-$7007& 0.1381   & 0.81 & 0.6   & 0.10 & $-$21.3 \\
SGRS~J1259$-$7737&  $>0.3$   & $>1.5$  & $>0.09$ & $>1.25$ & $<-18.5$\tablenotemark{b}  \\
SGRS~J1336$-$8019& 0.2478   & 2.34 & 1.57  & 0.88 & $-$23.7:\tablenotemark{c} \\
SGRS~J1728$-$7237& 0.4735   & 2.2  & 9.95  & 1.79 & $\le -$21.1\tablenotemark{d} \\
SGRS~J1911$-$7048& 0.2152   & 1.24 & 0.16  & 0.32 & $-$20.7  \\
SGRS~J1919$-$7959& 0.3462    & 1.75 & 1.03  & 5.00 & $\le -$19.3\tablenotemark{e} \\
SGRS~J1946$-$8222& 0.333:     & 2.14 & 0.76  & 0.80 & $\le -$18.9\tablenotemark{e}  \\
SGRS~J2159$-$7219& 0.0974   & 0.98 & 0.22  & 0.03 & $-$21.6 \\
\enddata
\tablenotetext{a}{The core luminosity given for SGRS~J0326$-$7730 and J0810$-$6800 
is using 4.8-GHz core flux and assuming flat core spectral index.}
\tablenotetext{b}{Redshift lower limit is based on K=16.2$\pm0.2$ mag and using K-z 
relation for radio galaxies. The core luminosity is using 4.8-GHz core flux and assuming 
flat core spectral index. The host galaxy is 
obscured; this is reflected in its absolute magnitude.}
\tablenotetext{c}{The host galaxy is in a close pair in which the objects appear equally bright; 
the object to the north 
is a star.  The galaxy magnitude taken from Table 3 has been estimated by dividing the 
integrated magnitude equally between the two objects.}
\tablenotetext{d}{The host galaxy of SGRS~J1728$-$7237 is partly obscured by a
star. The absolute magnitude is based on apparent magnitude judged from the 
neighbourhood galaxies 
within $5\arcsec$. The galaxy is likely to be more luminous.} 
\tablenotetext{e}{The host galaxies for SGRS~J1919$-$7959 and J1946$-$8222 are 
obscured. This is reflected in their absolute magnitudes.}
\end{deluxetable}

\clearpage

\begin{figure}
\epsscale{0.3}
\plotone{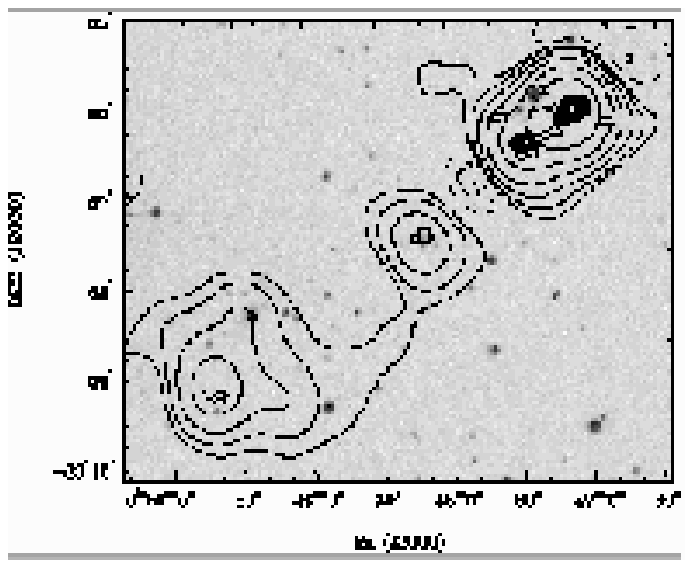}
\caption{SGRS~J0047$-$8307.  Radio-optical overlays for the SUMSS giant radio 
sources are shown in this and the following 17 figures. 
843~MHz contours for the low resolution SUMSS images are shown
in thick lines and 1.4-GHz (4.8~GHz where specified) contours for the 
higher resolution ATCA images are shown using thin lines. 
The radio contours are overlaid on the red SSS optical field shown using grey scales. 
Contours are at ($-$2, $-$1, 1, 2, 4, 8, 16 32, 64) $\times$ 4~mJy~beam$^{-1}$ 
for SUMSS and at ($-$2, $-$1, 1, 2, 3, 
4, 6, 8, 12, 16, 24) $\times$ 3~mJy~beam$^{-1}$ for ATCA. The beams have FWHM
$45\farcs 2$~$\times$~$45\arcsec$ (SUMSS) and $8\farcs 6$~$\times$~$5\farcs 9$ (ATCA).}
\end{figure}

\begin{figure}
(a) \hfill\break
\epsscale{0.3}
\plotone{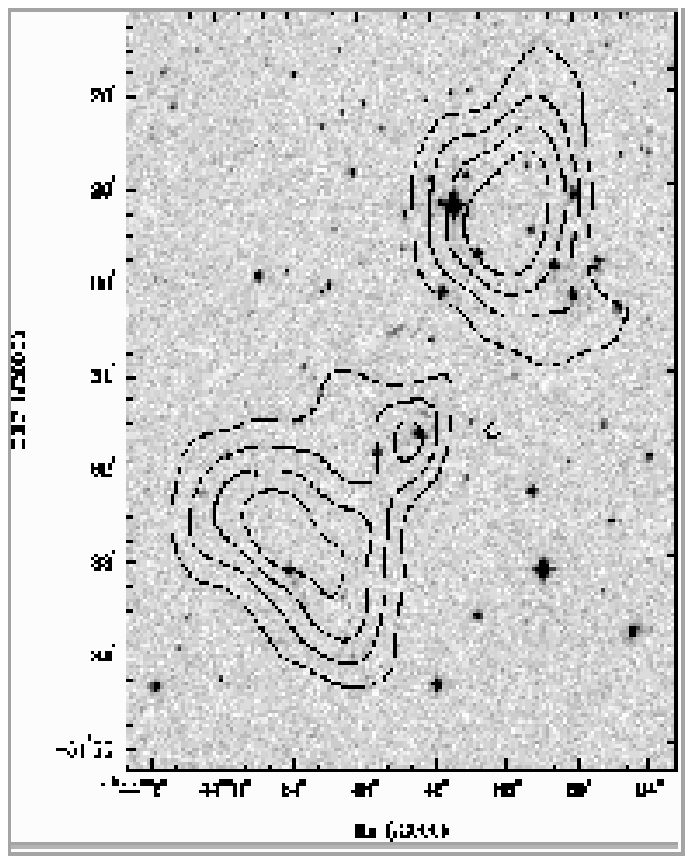}
(b) \hfill\break
\epsscale{0.3}
\plotone{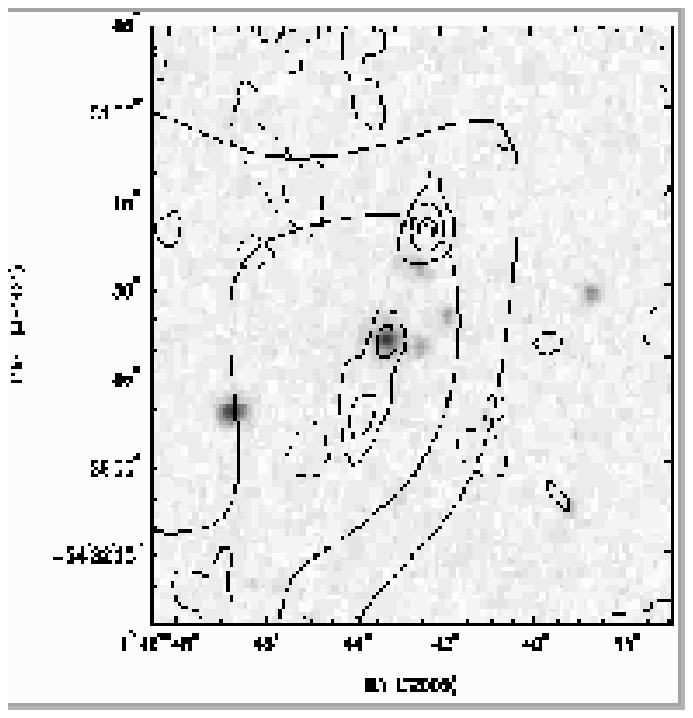}
\caption{SGRS~J0143$-$5431 radio contours on blue SSS grey scale.  SUMSS contours (thick lines)
at ($-$2, $-$1, 1, 2, 3, 4) $\times$ 4~mJy~beam$^{-1}$
and ATCA contours (thin lines) at ($-$2, $-$1, 1, 2, 3) $\times$ 0.7~mJy~beam$^{-1}$. 
Panel (b) shows a zoom of the region close to the core. The beams have FWHM
$54\farcs 3$~$\times$~$45\arcsec$ (SUMSS) and $8\farcs 5$~$\times$~$7\farcs 2$ (ATCA).}
\end{figure} 

\clearpage

\begin{figure}
\epsscale{0.3}
\plotone{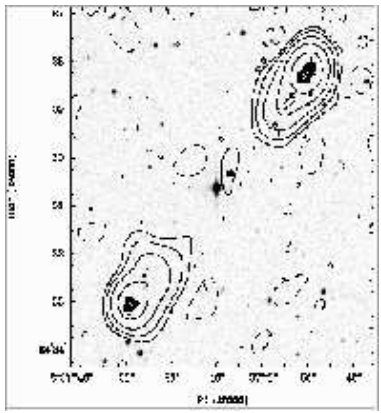}
\caption{SGRS~J0237$-$6429 radio contours on blue SSS grey scale. SUMSS contours (thick lines) at
($-$2, $-$1, 1, 2, 4, 8, 16) $\times$ 2~mJy~beam$^{-1}$ and ATCA contours (thin lines) at 
($-$2, $-$1, 1, 2, 3, 4, 6, 8,
12, 16, 24) $\times$ 2~mJy~beam$^{-1}$. The beams have FWHM
$50\farcs 1$~$\times$~$45\arcsec$ (SUMSS) and $7\farcs 6$~$\times$~$7\farcs 3$ (ATCA).} 
\end{figure}

\begin{figure}
(a) \hfill\break
\epsscale{0.3}
\plotone{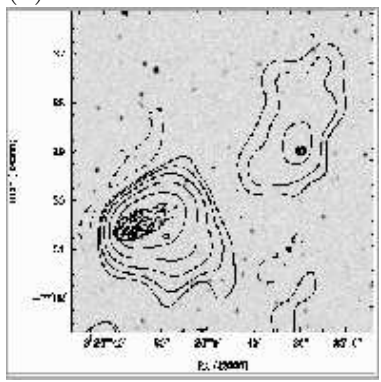}
(b) \hfill\break
\epsscale{0.3}
\plotone{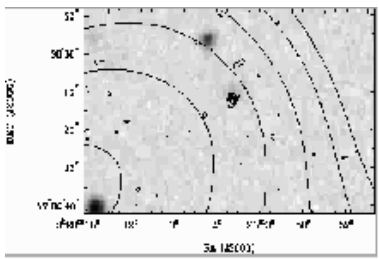}
\caption{SGRS~J0326$-$7730 radio contours on red SSS grey scale. 
A zoom of the core region is in panel (b) in which 4.8-GHz
ATCA contours are shown in thin lines. SUMSS contours (thick lines) are at  
($-$2, $-$1, 1, 2, 4, 8, 16, 32) $\times$ 3~mJy~beam$^{-1}$ and ATCA contours (thin lines) are at 
($-$2, $-$1, 1, 2, 3, 4, 6) $\times$ 0.7~mJy~beam$^{-1}$ for the 1.4-GHz image and at 
($-$2, $-$1, 1, 2, 3, 4) $\times$ 0.1~mJy~beam$^{-1}$ for the 4.8-GHz image. 
The beams have FWHM $46\farcs 4$~$\times$~$45\arcsec$ (SUMSS),  
$7\farcs 8$~$\times$~$4\farcs 9$ (ATCA, 1.4~GHz) 
and $2\farcs 2$~$\times$~$1\farcs 8$ (ATCA, 4.8~GHz).}
\end{figure}

\clearpage

\begin{figure}
\epsscale{0.1}
(a) \hfill\break
\includegraphics[width=40mm]{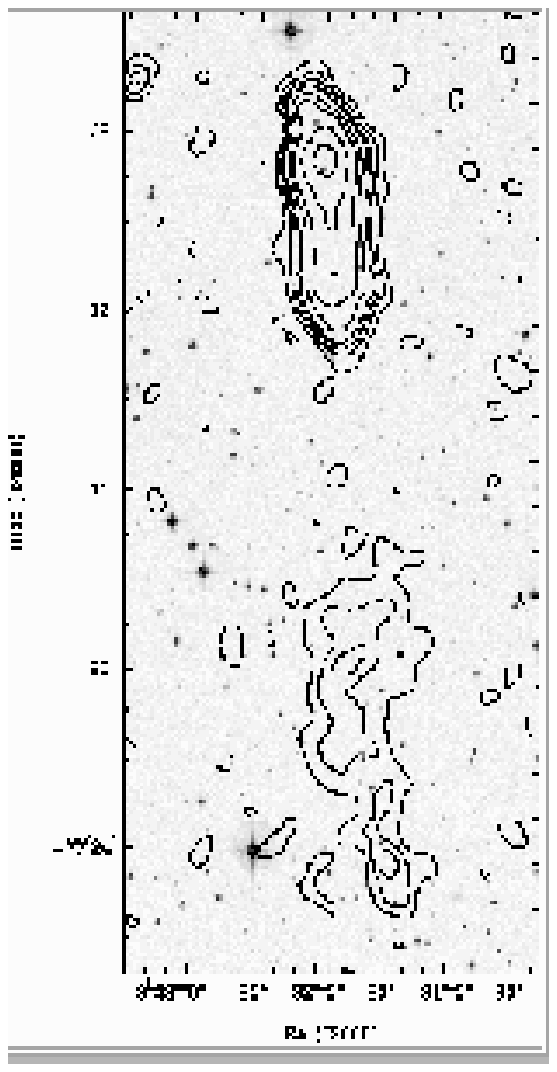} \hfill\break
(b) \hfill\break
\epsscale{0.1}
\includegraphics[width=40mm]{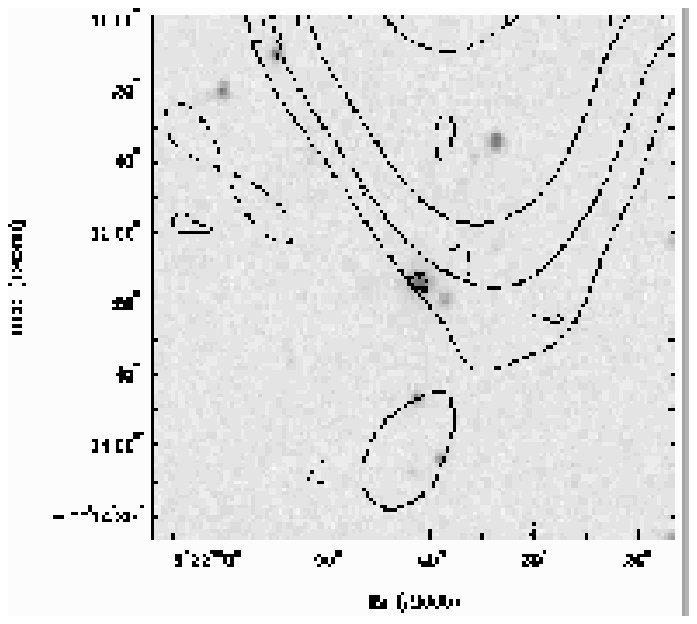}
\caption{SGRS~J0331$-$7710 radio contours on blue SSS grey scale. 
A zoom of the core region is in panel (b). SUMSS contours (thick lines) are at ($-$2, $-$1, 1, 2, 3, 4, 6, 8, 12, 16) $\times$ 2~mJy~beam$^{-1}$ in panel (a) and at 
($-$1, 1, 2, 4, 8) $\times$ 3~mJy~beam$^{-1}$ in panel (b); ATCA contours (thin lines) are at
($-$1, 1, 1.5) $\times$ 0.8~mJy~beam$^{-1}$. The beams 
have FWHM $46\farcs 4$~$\times$~$45\arcsec$ (SUMSS) and 
$9\farcs 1$~$\times$~$6\arcsec$ (ATCA).}
\end{figure}

\begin{figure}
(a) \hfill\break
\epsscale{0.3}
\plotone{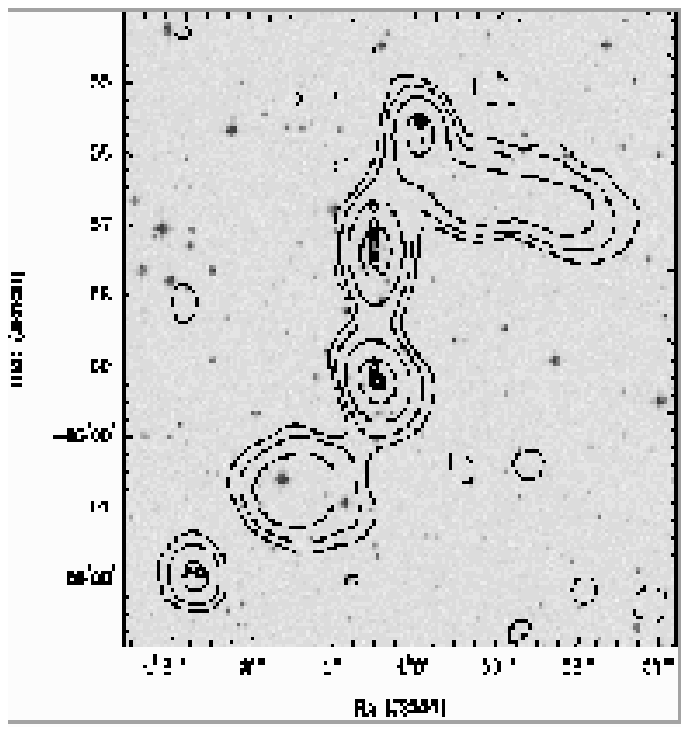}
(b) \hfill\break
\epsscale{0.3}
\plotone{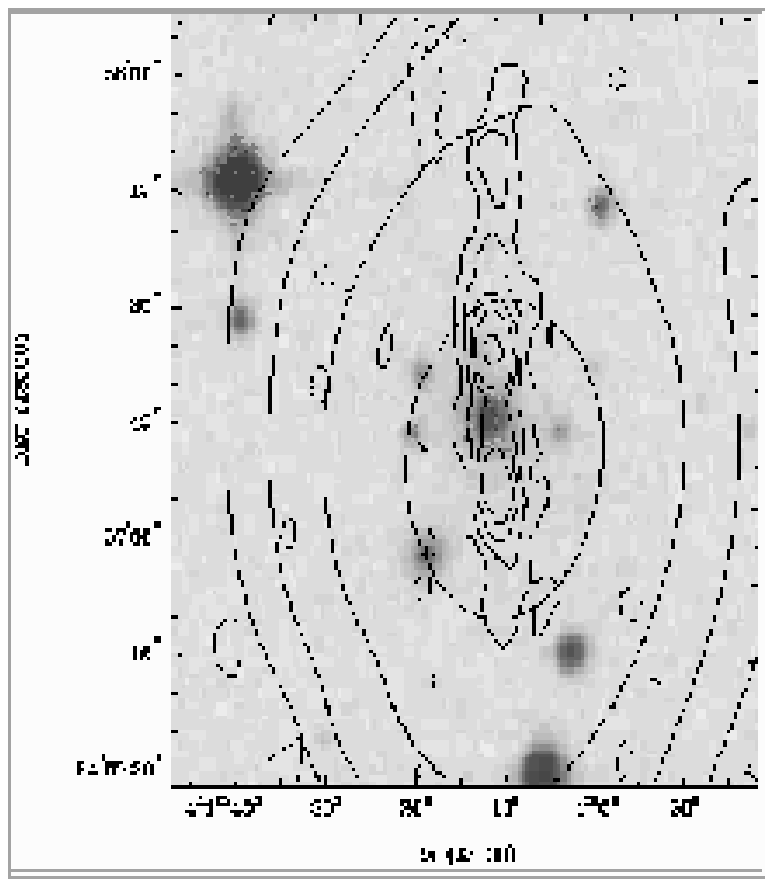}
\end{figure}

\clearpage

\begin{figure}
(c) \hfill\break
\epsscale{0.3}
\plotone{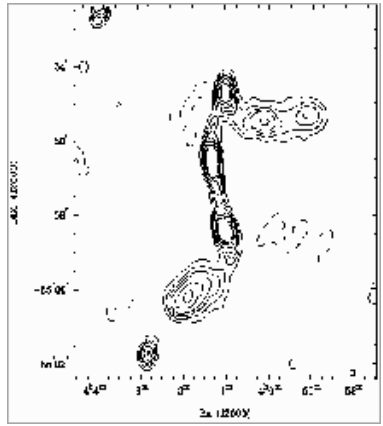}
\caption{SGRS~J0400$-$8456 radio contours on blue SSS grey scale. 
A zoom of the core region is in panel (b).
SUMSS contours (thick lines) are at  
($-$1, 1, 2, 4, 8) $\times$ 2~mJy~beam$^{-1}$ in panel (a) and at
($-$1, 1, 2, 4, 8) $\times$ 3~mJy~beam$^{-1}$ in panel (b); ATCA contours (thin lines) are at 
($-$1, 1, 2, 3, 4) $\times$ 1~mJy~beam$^{-1}$ in panel (a) and at ($-$1, 1, 2, 3, 4, 6) 
$\times$ 0.1~mJy~beam$^{-1}$ in panel (b). The beams have FWHM 
$45\farcs 2$~$\times$~$45\arcsec$ (SUMSS) and $8\farcs 1$~$\times$~$4\farcs 7$ (ATCA).Low-resolution 1.4-GHz ATCA image of SGRS~J0400$-$8456. 
The contours are at ($-$2, $-$1, 1, 2, 3, 4, 6, 8, 10)$\times$ 0.8~mJy~beam$^{-1}$. 
The beam FWHM is $34\arcsec \times 21\arcsec$.} 
\end{figure}

\begin{figure}
(a) \hfill\break
\epsscale{0.3}
\plotone{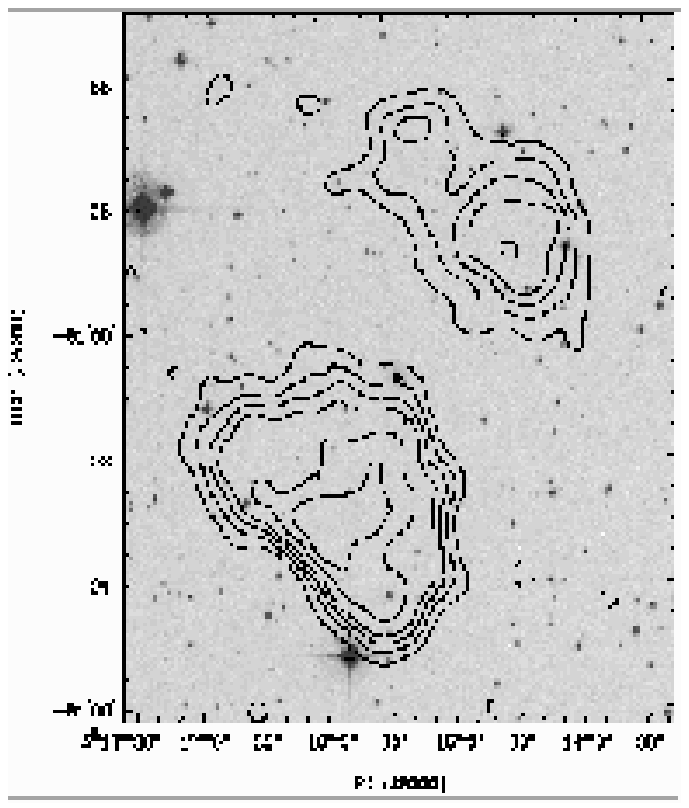}
(b) \hfill\break
\epsscale{0.3}
\plotone{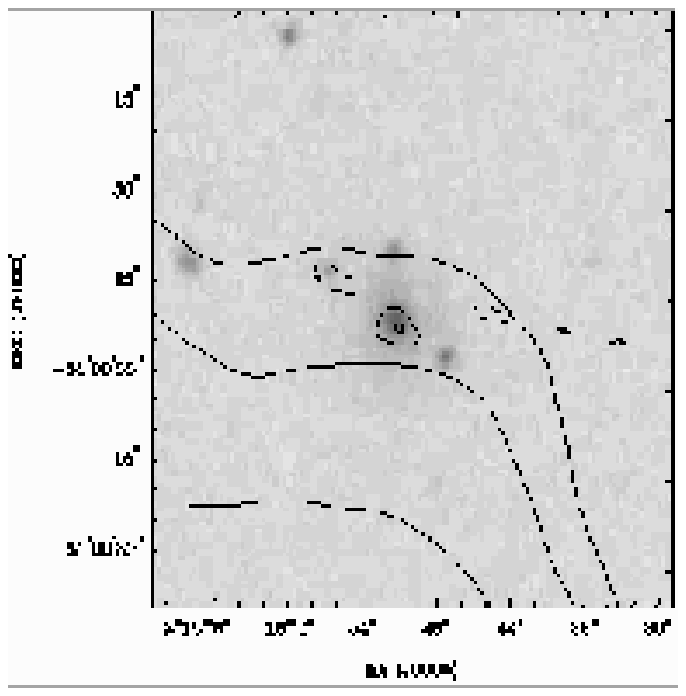}
\caption{SGRS~J0515$-$8100  radio contours on blue SSS grey scale. 
A zoom of the core region is in panel (b).
SUMSS contours (thick lines) are at ($-$1, 1, 2, 3, 4, 6, 8) $\times$ 1.5~mJy~beam$^{-1}$,
ATCA contours (thin lines) are at 2~mJy~beam$^{-1}$ in panel (a) and at
($-$1, 1, 2) $\times$ 1~mJy~beam$^{-1}$ in panel (b). The beams have FWHM  
$45\farcs 7$~$\times$~$45\arcsec$ (SUMSS), $8\farcs 5$~$\times$~$6\farcs 1$ 
(ATCA image in panel (a)) and
$7\farcs 1$~$\times$~$5\arcsec$ (ATCA image in panel (b)).} 
\end{figure}

\clearpage

\begin{figure}
\epsscale{0.3}
\plotone{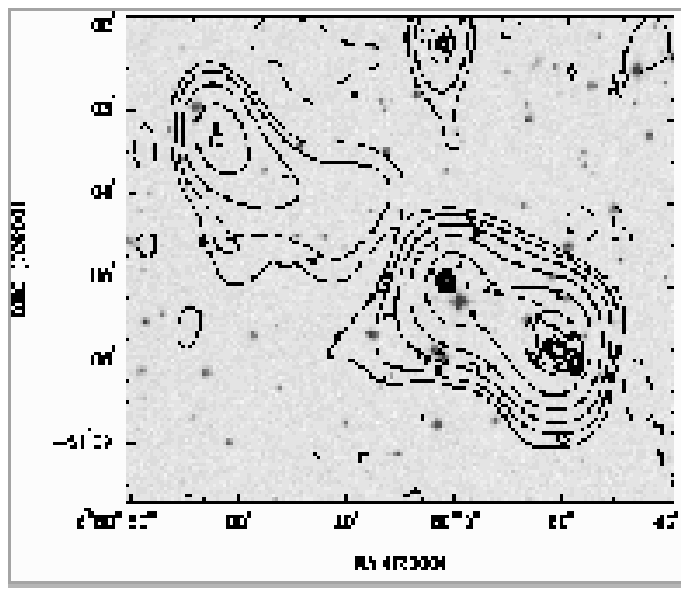}
\caption{SGRS~J0631$-$5405 radio contours on blue SSS grey scale. 
SUMSS contours (thick lines) are at ($-$1, 1, 2, 4, 8, 16, 32) 
$\times$ 4~mJy~beam$^{-1}$ and ATCA contours (thin lines) are at
($-$1, 1, 2, 3, 4, 6, 8, 12) $\times$ 2.5~mJy~beam$^{-1}$. The beams have FWHM
$54\farcs 3$~$\times$~$45\arcsec$ (SUMSS) and $8\farcs 5$~$\times$~$7\farcs 3$ (ATCA).} 
\end{figure}

\begin{figure}
\epsscale{0.3}
\plotone{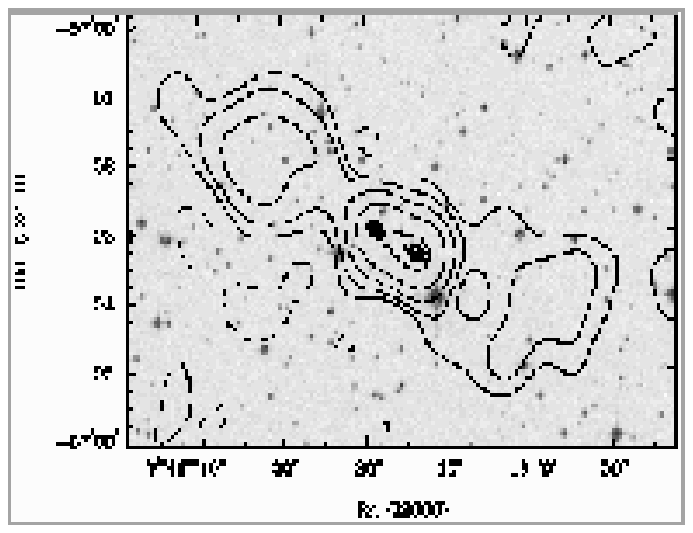}
\caption{SGRS~J0746$-$5702 radio contours on blue SSS grey scale. 
SUMSS contours (thick lines) are at ($-$1, 1, 2, 4, 8, 16) 
$\times$ 3~mJy~beam$^{-1}$ and ATCA contours (thin lines) are at
($-$1, 1, 2, 3, 4, 6, 8, 12) $\times$ 3~mJy~beam$^{-1}$. The beams have FWHM  
$54\farcs 3$~$\times$~$45\arcsec$ (SUMSS) and $8\farcs 4$~$\times$~$7\farcs 1$ (ATCA).} 
\end{figure}

\begin{figure}
(a) \hfill\break
\epsscale{0.3}
\plotone{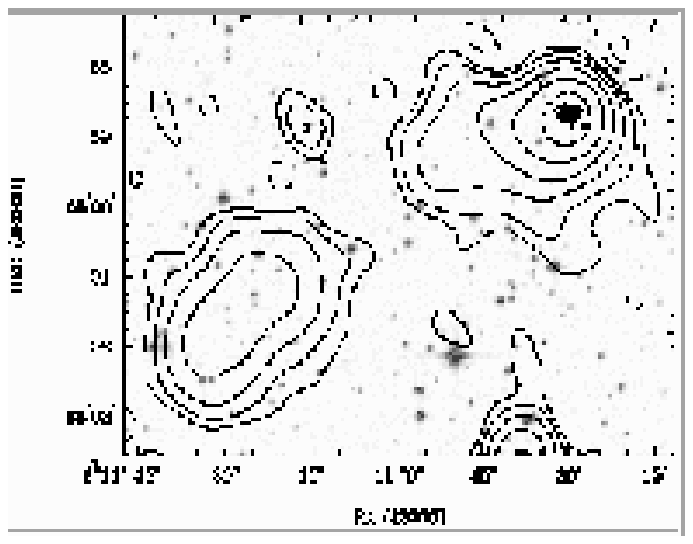}
(b) \hfill\break
\epsscale{0.3}
\plotone{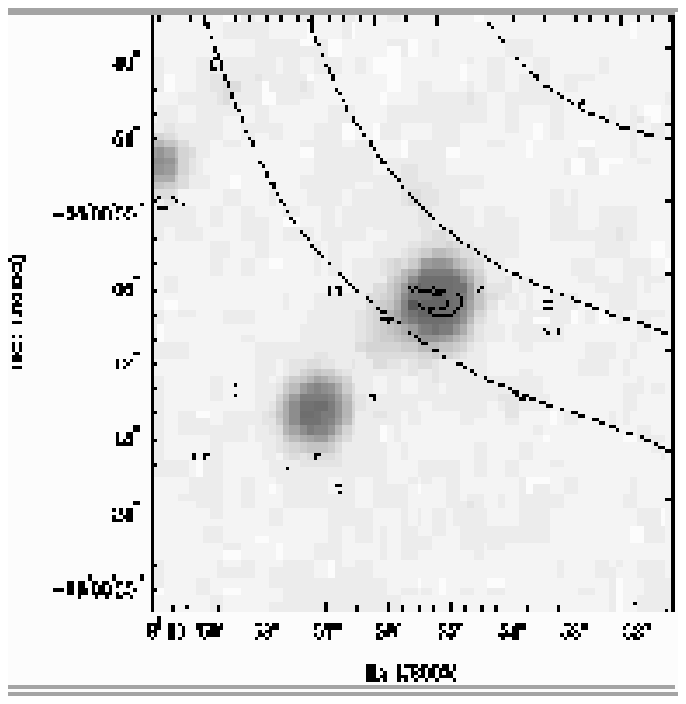}
\caption{SGRS~J0810$-$6800 radio contours on red SSS grey scale.  The core region is shown in panel (b) with 
4.8-GHz contours plotted in thin lines. SUMSS contours (thick lines) are at ($-$1, 1, 2, 4, 8, 16, 32) 
$\times$ 1.5~mJy~beam$^{-1}$.  ATCA contours (thin lines) are at 
($-$1, 1, 2, 3, 4, 6, 8, 12) $\times$ 1.5~mJy~beam$^{-1}$ for the 1.4-GHz image and at
($-$1, 1, 2) $\times$ 0.1~mJy~beam$^{-1}$ for the 4.8-GHz image. The beams have FWHM 
$48\farcs 5$~$\times$~$45\arcsec$ (SUMSS), $9\farcs 3$~$\times$~$6\farcs 1$ 
(ATCA 1.4~GHz) and $3\farcs 4$~$\times$~$1\farcs 4$ (ATCA 4.8~GHz).}
\end{figure}

\clearpage

\begin{figure}
\epsscale{0.3}
\plotone{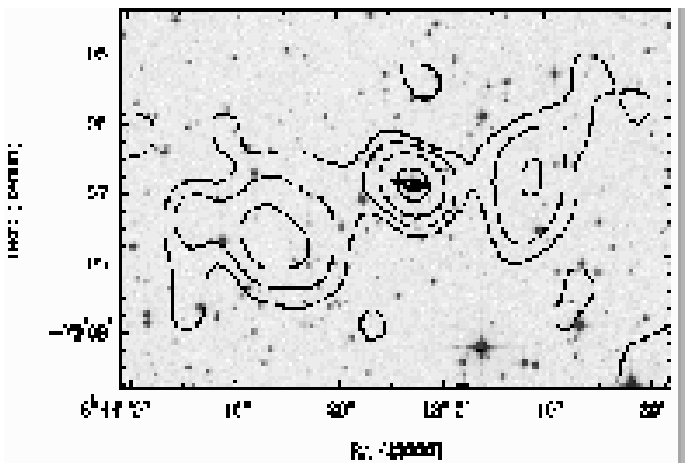}
\caption{SGRS~J0843$-$7007 radio contours on blue SSS grey scale. 
SUMSS contours (thick lines) are at ($-$1, 1, 2, 4, 8) 
$\times$ 3~mJy~beam$^{-1}$ and ATCA contours (thin lines) are at
($-$1, 1, 2, 3, 4) $\times$ 3~mJy~beam$^{-1}$. The beams have FWHM  
$47\farcs 3$~$\times$~$45\arcsec$ (SUMSS) and $9\farcs 5$~$\times$~$6\arcsec$ (ATCA).}
\end{figure}

\begin{figure}
(a) \hfill\break
\epsscale{0.3}
\plotone{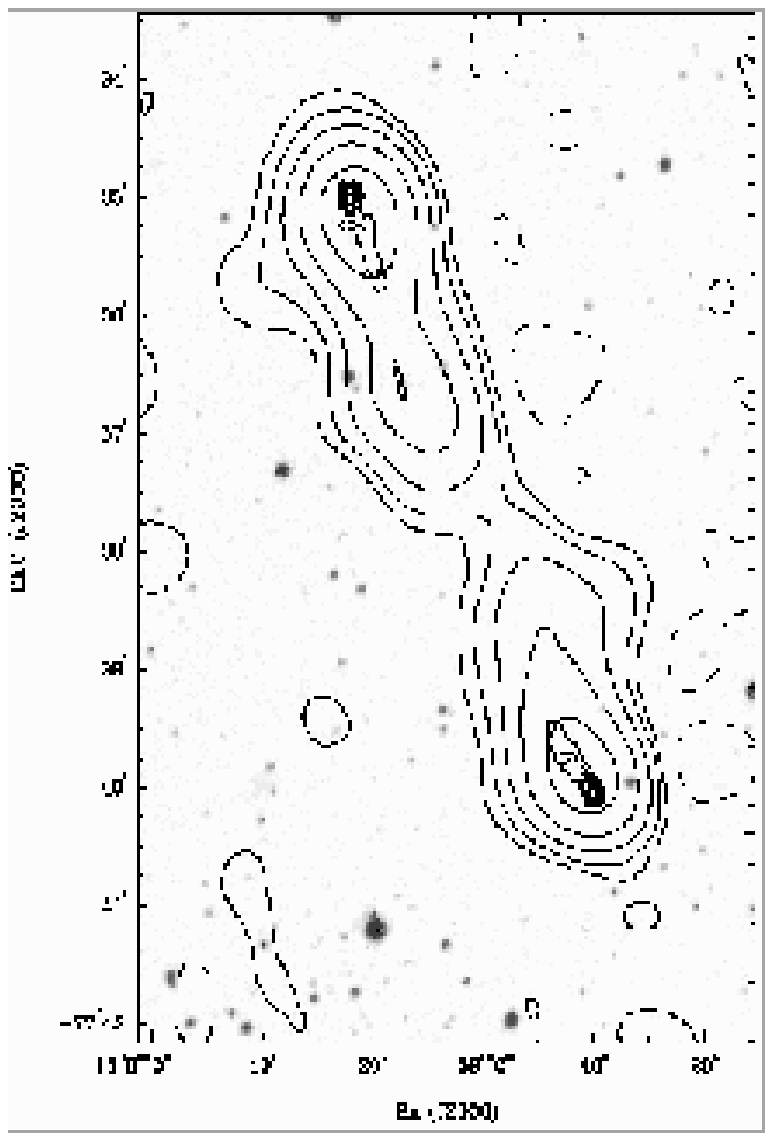}
(b) \hfill\break
\epsscale{0.3}
\plotone{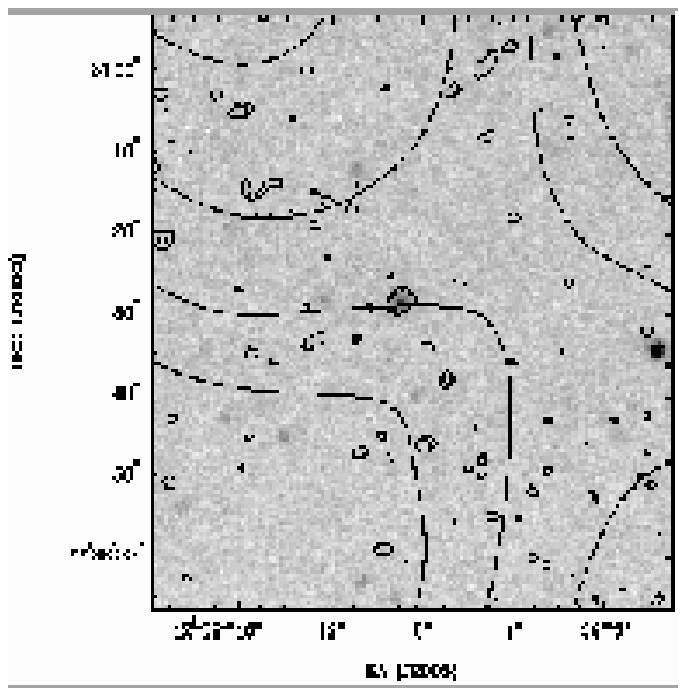}
\caption{SGRS~J1259$-$7737 radio contours.  Blue SSS grey scale image is in panel (a) 
and SSO 2.3 m K-band image in grey scale is in panel (b). The K-band image was 
obtained by Helen Johnston. The core region is shown in panel (b) with 
4.8-GHz contours shown using thin lines.  SUMSS contours (thick lines) are at ($-$1, 1, 2, 4, 8, 16) 
$\times$ 4~mJy~beam$^{-1}$, ATCA contours (thin lines) are at 
($-$1, 1, 2, 3, 4, 6, 8) $\times$ 3~mJy~beam$^{-1}$ for the 1.4-GHz image and at
($-$1, 1, 2) $\times$ 0.07~mJy~beam$^{-1}$ for the 4.8-GHz image. 
The beams have FWHM  
$46\farcs 4$~$\times$~$45\arcsec$ (SUMSS), $9\farcs 4$~$\times$~$5\farcs 8$ (ATCA 1.4~GHz) and
$2\farcs 2$~$\times$~$1\farcs 8$ (ATCA 4.8~GHz).}
\end{figure}

\clearpage

\begin{figure}
(a) \hfill\break
\epsscale{0.1}
\includegraphics[width=40mm]{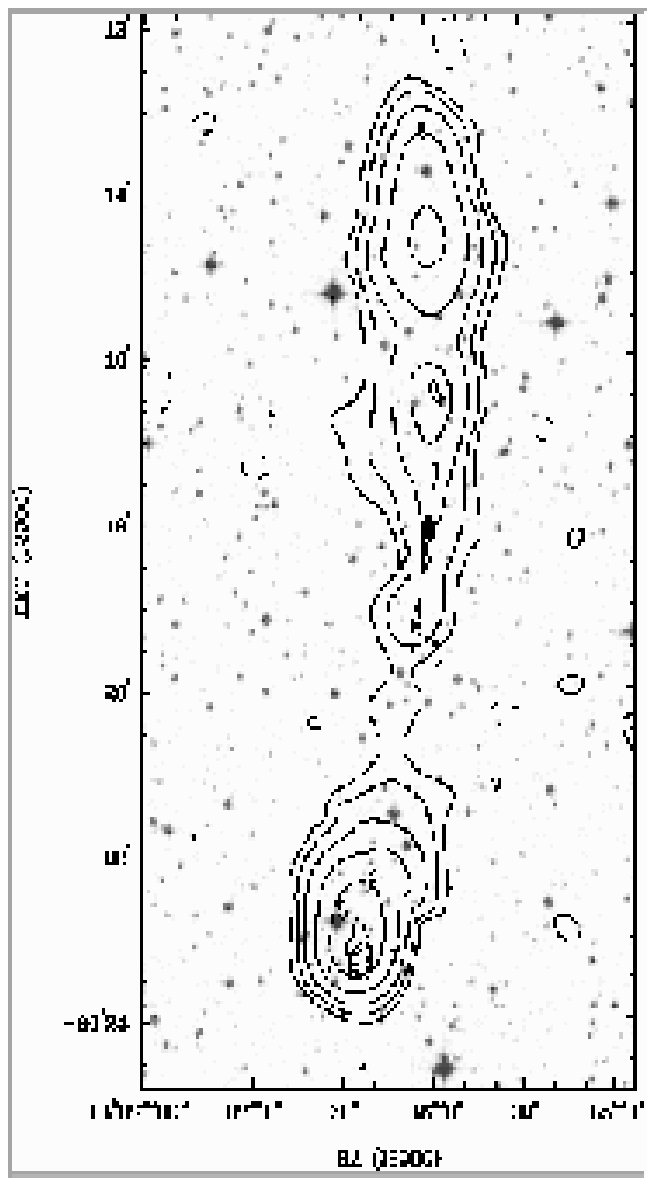} \hfill\break
(b) \hfill\break
\epsscale{0.1}
\includegraphics[width=40mm]{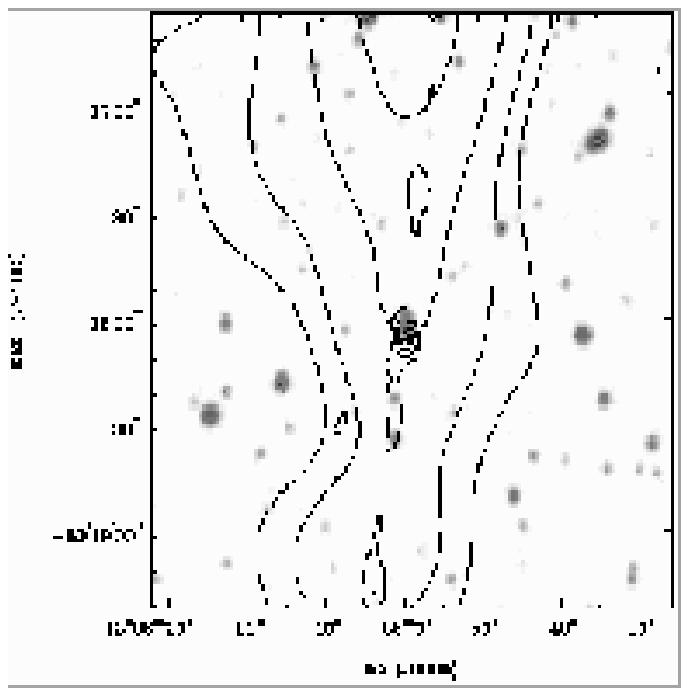}
\caption{SGRS~J1336$-$8019 radio contours on blue SSS grey scale. 
A zoom of the core region is in panel (b).
SUMSS contours (thick lines) are at ($-$1, 1, 2, 4, 8, 16) $\times$ 3~mJy~beam$^{-1}$ and 
ATCA contours (thin lines) are at ($-$1, 1, 2, 3, 4) $\times$ 2~mJy~beam$^{-1}$. The beams have a FWHM 
$45\farcs 7$~$\times$~$45\arcsec$ (SUMSS) and $9\farcs 3$~$\times$~$5\farcs 8$ (ATCA).}
\end{figure}

\begin{figure}
\epsscale{0.3}
\plotone{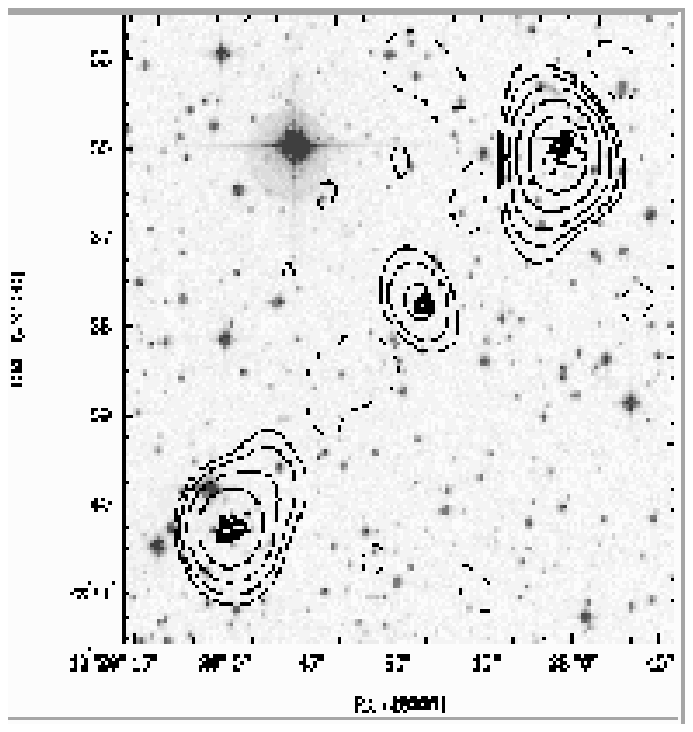}
\caption{SGRS~J1728$-$7237 radio contours on blue SSS grey scale. 
SUMSS contours (thick lines) are at ($-$1, 1, 2, 4, 8, 16) 
$\times$ 3~mJy~beam$^{-1}$ and ATCA contours (thin lines) are at
($-$1, 1, 2, 3, 4, 6, 8, 12) $\times$ 2.5~mJy~beam$^{-1}$. The beams have FWHM 
$47\farcs 3$~$\times$~$45\arcsec$ (SUMSS) and $8\farcs 9$~$\times$~$6\farcs 1$ (ATCA).} 
\end{figure}

\clearpage

\begin{figure}
\epsscale{0.3}
\plotone{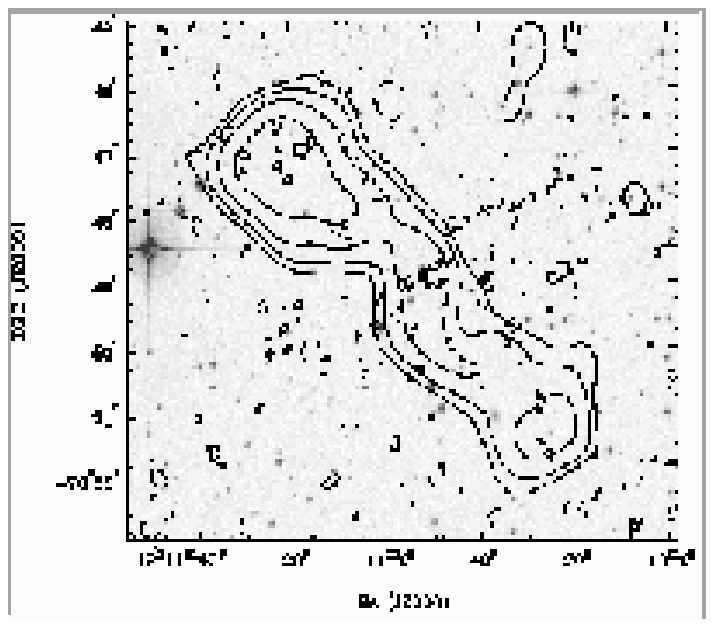}
\caption{SGRS~J1911$-$7048 radio contours on blue SSS grey scale. 
SUMSS contours (thick lines) are at ($-$1, 1, 2, 4, 8) 
$\times$ 2~mJy~beam$^{-1}$ and ATCA contours (thin lines) are at
($-$1, 1, 2) $\times$ 0.5~mJy~beam$^{-1}$. The beams have FWHM 
$47\farcs 3$~$\times$~$45\arcsec$ (SUMSS) and $9\farcs 8$~$\times$~$5\farcs 9$ (ATCA).}
\end{figure}

\begin{figure}
\epsscale{0.3}
\plotone{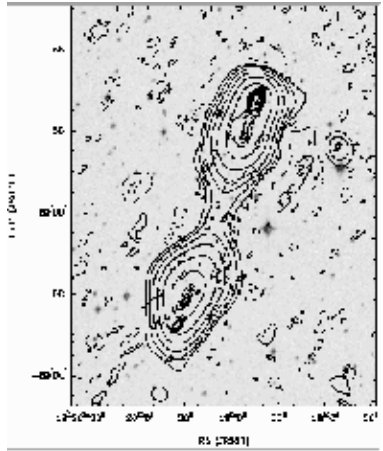}
\caption{SGRS~J1919$-$7959 radio contours on red SSS grey scale. 
SUMSS contours (thick lines) are at ($-$1, 1, 2, 4, 8, 16, 32, 64) 
$\times$ 4~mJy~beam$^{-1}$ and ATCA contours (thin lines) are at 
($-$1, 1, 2, 3, 4, 6, 8, 12, 16) $\times$ 2~mJy~beam$^{-1}$. The beams have FWHM 
$45\farcs 7$~$\times$~$45\arcsec$ (SUMSS) and $11\farcs 8$~$\times$~$5\farcs 1$ (ATCA).}
\end{figure}

\begin{figure}
\epsscale{0.3}
\plotone{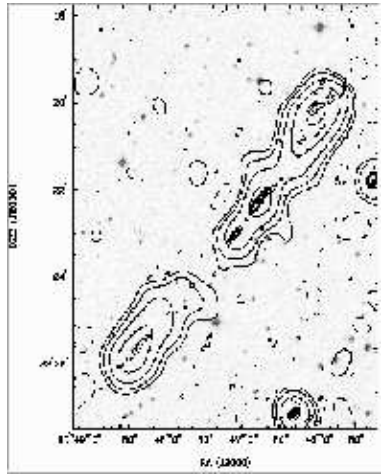}
\caption{SGRS~J1946$-$8222 radio contours on red SSS grey scale. 
SUMSS contours (thick lines) are at ($-$1, 1, 2, 4, 8, 16) 
$\times$ 2~mJy~beam$^{-1}$ and ATCA contours (thin lines) are at 
($-$1, 1, 2, 3, 4, 6, 8, 12, 16) $\times$ 0.6~mJy~beam$^{-1}$. The beams have FWHM 
$45\farcs 2$~$\times$~$45\arcsec$ (SUMSS) and $10\farcs 2$~$\times$~$4\farcs 3$ (ATCA).}
\end{figure}

\clearpage

\begin{figure}
(a) \hfill\break
\epsscale{0.3}
\plotone{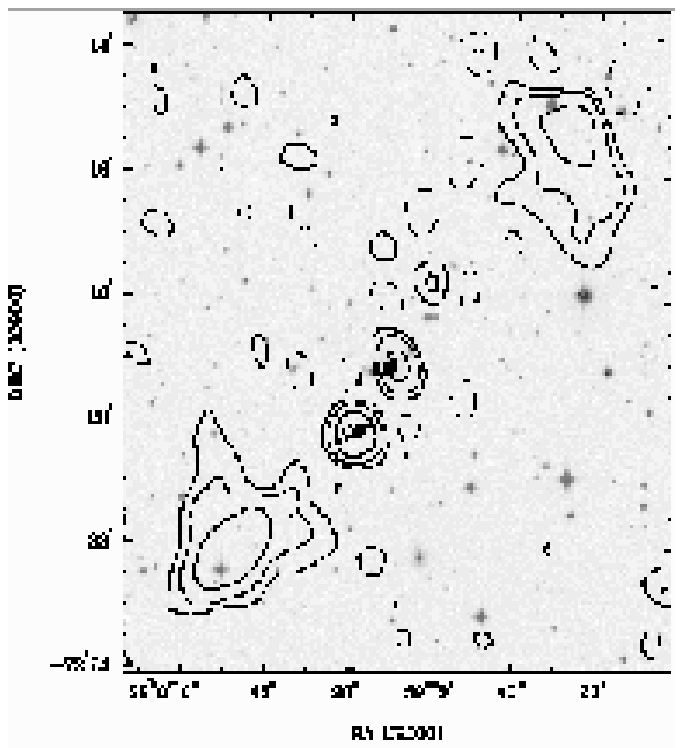}
(b) \hfill\break
\plotone{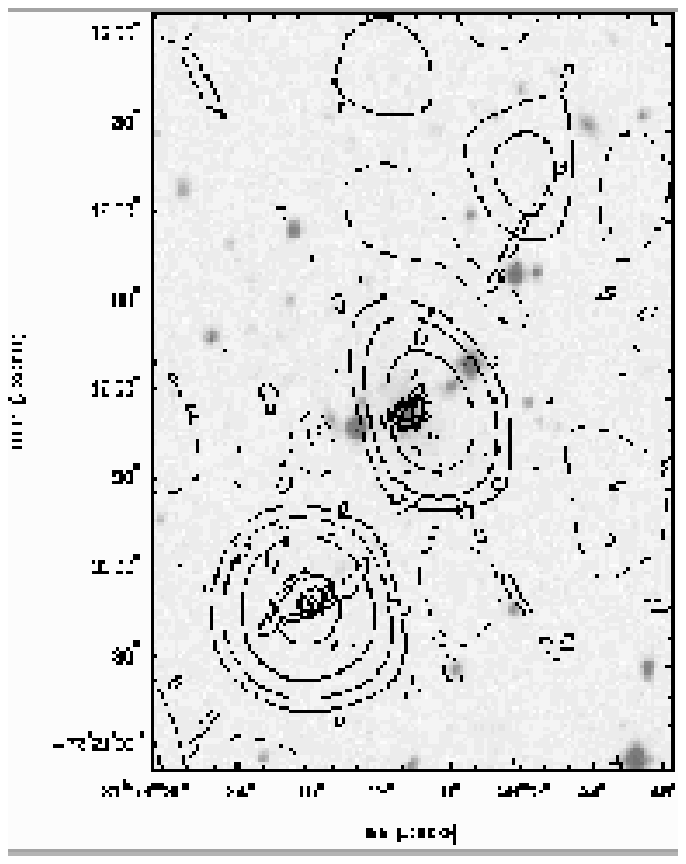}
\epsscale{0.3}
\caption{SGRS~J2159$-$7219 radio contours on blue SSS grey scale. A zoom of the core region is shown in 
panel (b). SUMSS contours (thick lines) are at  ($-$1, 1, 2, 4) $\times$ 2.5~mJy~beam$^{-1}$ 
and ATCA contours (thin lines) are at ($-$1, 1, 2, 3, 4, 6) $\times$ 1.5~mJy~beam$^{-1}$ in panel (a), 
In panel (b), SUMSS contours are at ($-$1, 1, 2, 4, 8) $\times$ 2~mJy~beam$^{-1}$  and 
ATCA contours are at ($-$1, 1, 2, 3, 4, 6, 8) $\times$ 1~mJy~beam$^{-1}$. The beams have FWHM 
$47\farcs 3$~$\times$~$45\arcsec$ (SUMSS) and $9\farcs 4$~$\times$~$6\arcsec$ (ATCA).} 
\end{figure}

\clearpage
\onecolumn

\begin{figure*}
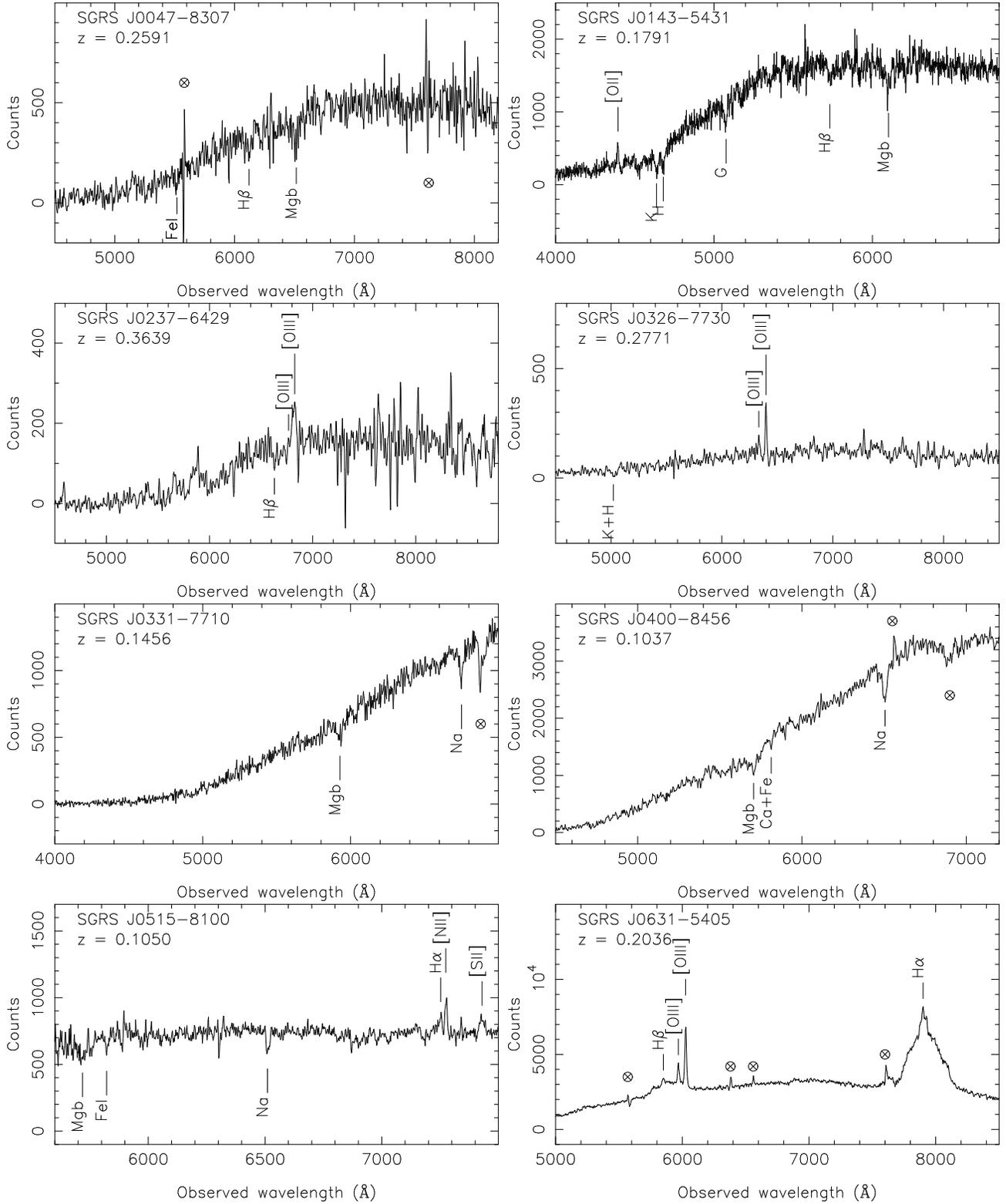

\centerline{
\includegraphics[height=85mm,angle=-90]{f19a.eps}
\includegraphics[height=85mm,angle=-90]{f19b.eps}
}
\centerline{
\includegraphics[height=85mm,angle=-90]{f19c.eps}
\includegraphics[height=85mm,angle=-90]{f19d.eps}
}
\centerline{
\includegraphics[height=85mm,angle=-90]{f19e.eps}
\includegraphics[height=85mm,angle=-90]{f19f.eps}
}
\centerline{
\includegraphics[height=85mm,angle=-90]{f19g.eps}
\includegraphics[height=85mm,angle=-90]{f19h.eps}
}
\caption{
Optical spectra of the complete sample of giant radio sources. The symbol $\otimes$ represents
features which we consider to be artifacts. 
}
\end{figure*}

\clearpage

\begin{figure*}
\centerline{
\includegraphics[height=85mm,angle=-90]{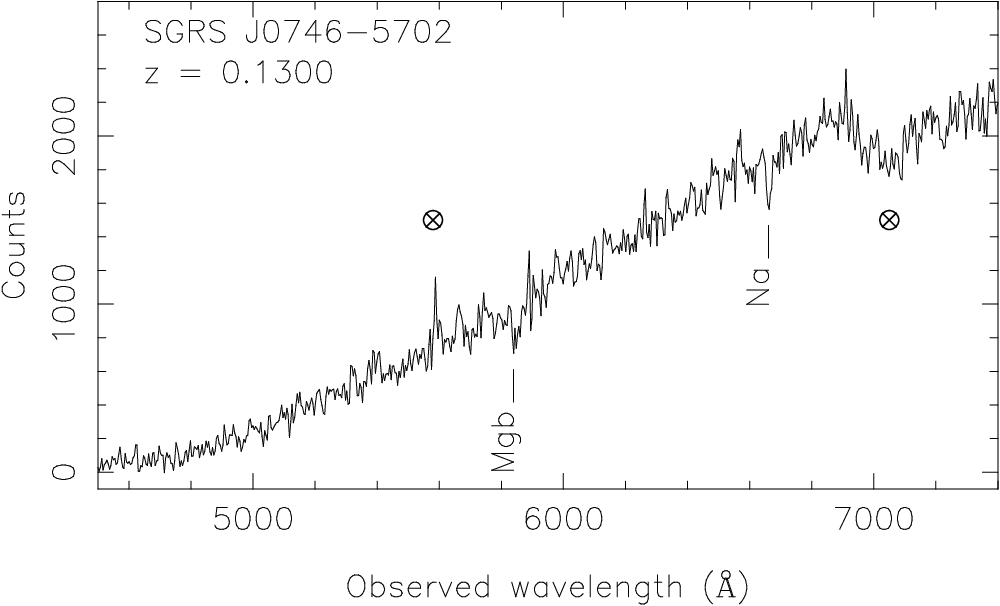}
\includegraphics[height=85mm,angle=-90]{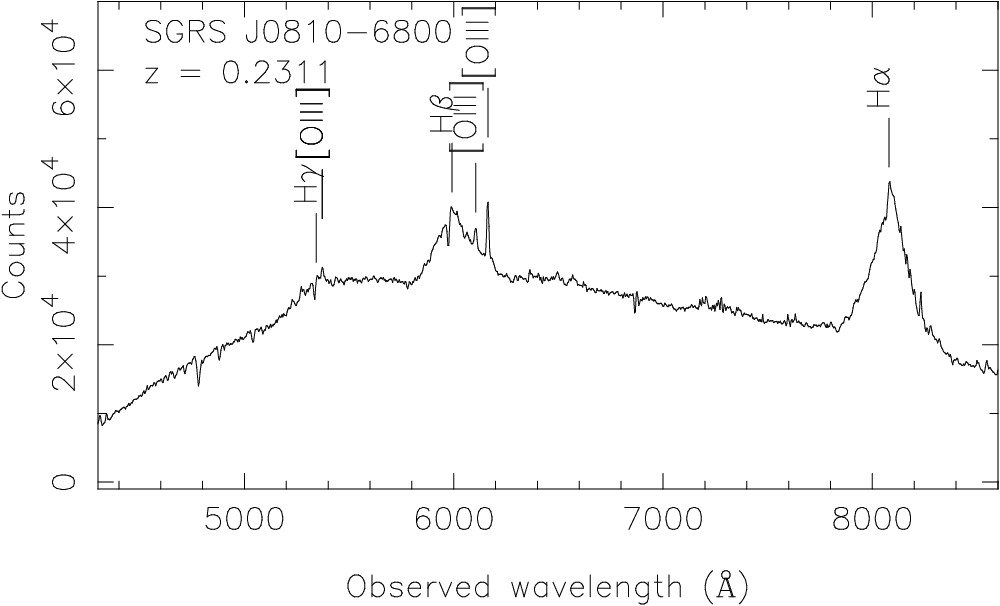}
}
\centerline{
\includegraphics[height=85mm,angle=-90]{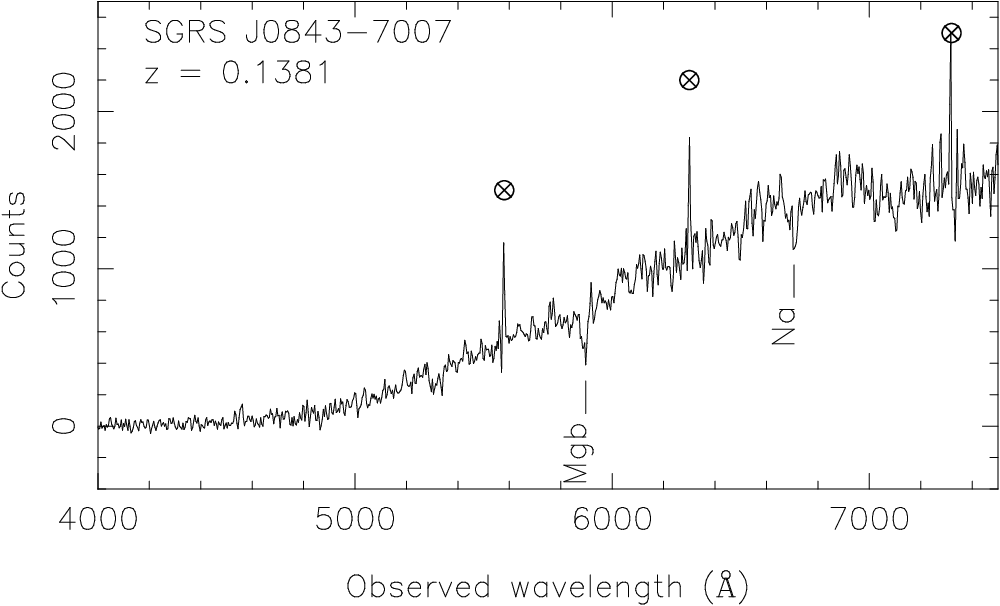}
\includegraphics[height=85mm,angle=-90]{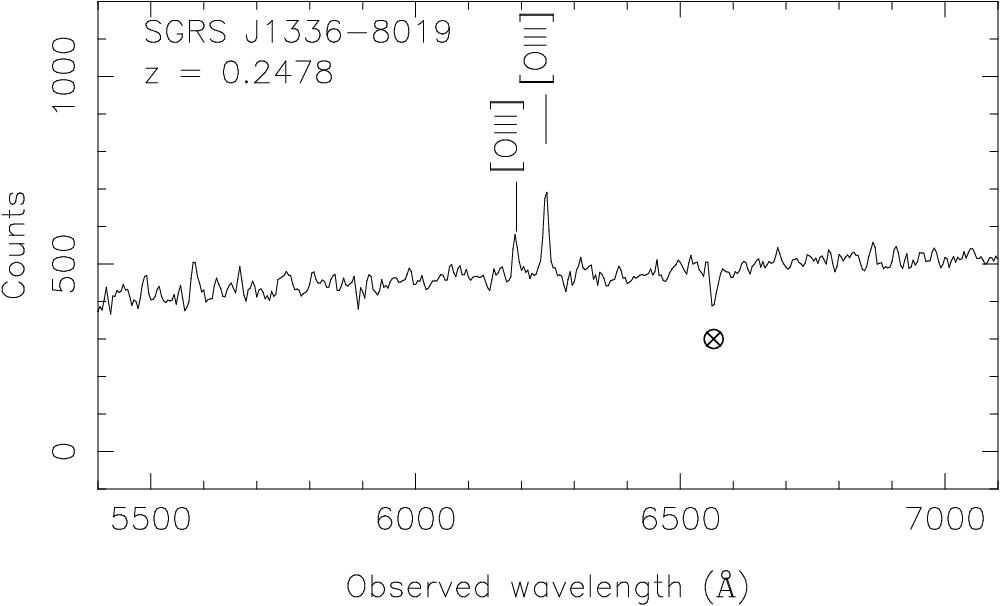}
}
\centerline{
\includegraphics[height=85mm,angle=-90]{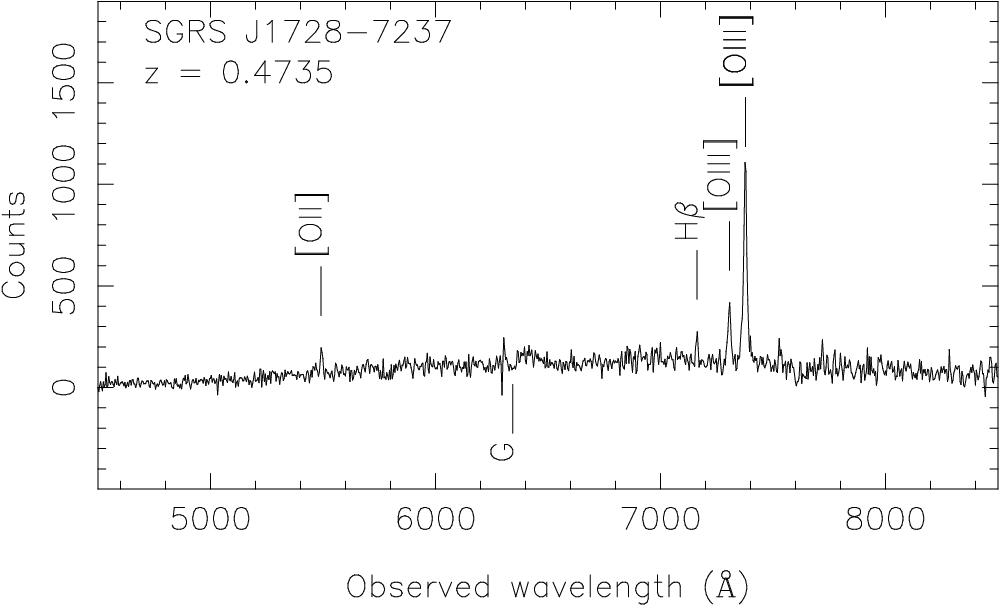}
\includegraphics[height=85mm,angle=-90]{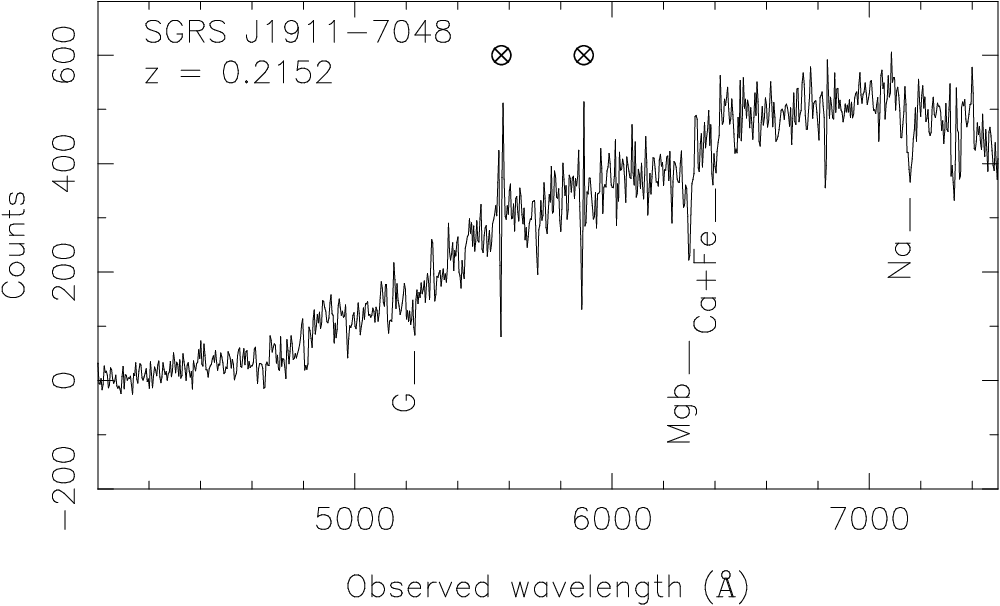}
}
\centerline{
\includegraphics[height=85mm,angle=-90]{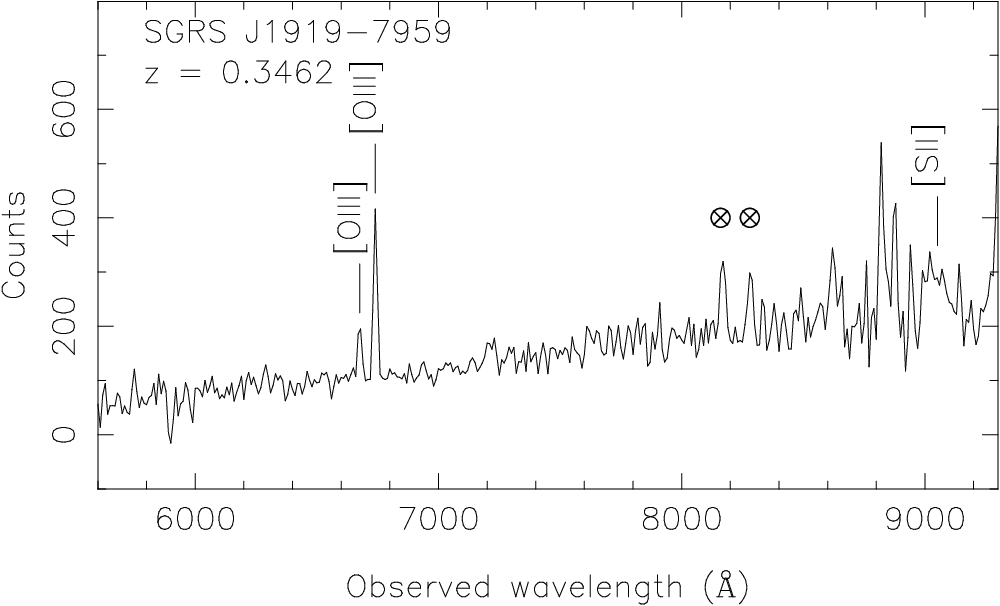}
\includegraphics[height=85mm,angle=-90]{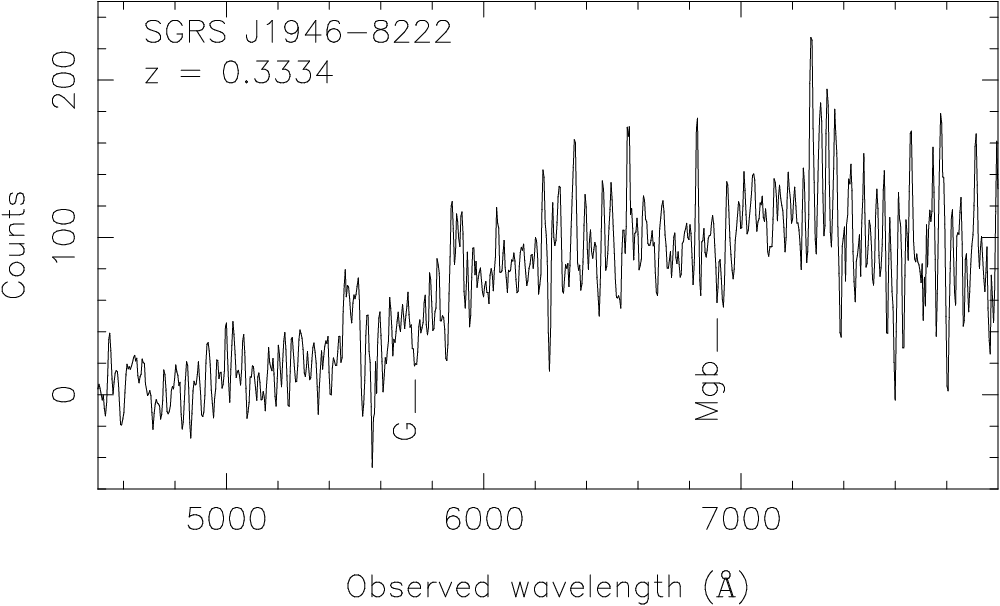}
}
\addtocounter{figure}{-1}
\caption{ continued.}
\end{figure*} 

\clearpage

\begin{figure*}
\centerline{
\includegraphics[height=85mm,angle=-90]{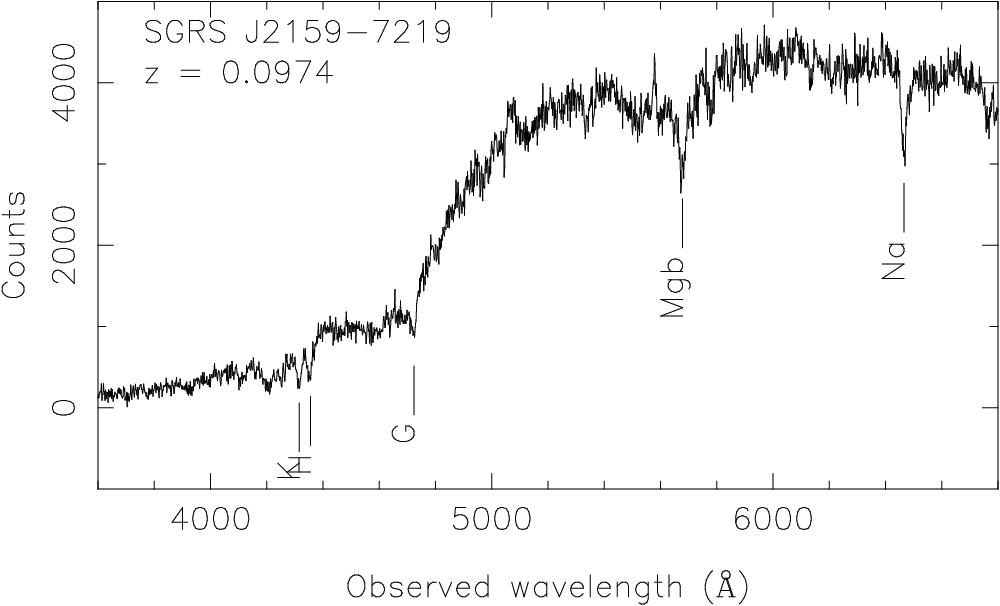}
}
\addtocounter{figure}{-1}
\caption{continued.}
\end{figure*}

\clearpage

\begin{deluxetable}{lcl}
\tablewidth{0pt}
\tablenum{A1}
\tablecaption{Redshifts and spectral lines in spectra in Fig.~29}
\startdata
\tableline
\tableline
Name          & Redshift   & Prominent lines    \\        
\tableline
SGRSC~J0020$-$7321    & 0.0839$\pm0.0001$     & Absorption lines: K, H, G, H$\beta$, Mgb, Na\\ 
SGRSC~J0129$-$6433    & 0.1343$\pm0.0002$     & Strong narrow emission lines: H$\beta$, [O{\sc iii}]4959,5007,\\
                    &            & H$\alpha$+[N{\sc ii}] blend and [S{\sc ii}]6716,6731\\      
                    &            & Weak emission lines: H$\gamma$+[O{\sc iii}]4363, He{\sc ii}4686\\ 
                    &            & and [OI]6300 \\
SGRSC~J0414$-$6933    & 0.228:     & Possible weak emission lines: H$\alpha$+[N{\sc ii}] blend and \\
                    &            & [S{\sc ii}]6716,6731. Absorption lines: Mgb and Na\\
SGRSC~J2228$-$5600    & 0.0796$\pm0.0002$     & Strong emission lines: [O{\sc iii}]4959,5007, [OI] 6300,\\
                    &            & H$\alpha$+[N{\sc ii}] blend, [S{\sc ii}]6716,6731 \\
                    &            & Absorption lines: Mgb and Na \\
SGRSC~J2336$-$8151    & 0.1190$\pm0.0001$     & Absorption lines: K, H, G, H$\beta$, Mgb and Na \\ 
\enddata
\end{deluxetable}

\clearpage

\begin{deluxetable}{lllcccc}
\rotate
\tabletypesize{\footnotesize}
\tablewidth{0pt}
\tablenum{A2}
\tablecaption{Parameters for some of the rejected candidates}
\startdata
\tableline
\tableline
Name  &     Other names & SUMSS   & Redshift & Angular size & 
Type            & Linear             \\
      &      & components &          & (arcmin)     & 
FR-{\sc i}/FR-{\sc ii}      & size (Mpc)         \\
\tableline
SGRSC~J0020$-$7321&PMN J0020$-$7321&J002046$-$732132& 0.0839   & 4.8                &   FR-{\sc ii}         & 0.45        \\ 
               &PMN J0021$-$7322&J002021$-$732115&          &                    &                 &           \\   
                      
               &              &J002124$-$732219&          &                    &                 &             \\
SGRSC~J0129$-$6433&PMN J0129$-$6433&J012934$-$643353& 0.1343   & 4.7                &   FR-{\sc ii}         & 0.66      \\
               &              &J012913$-$643223&          &                    &                 &                \\
               &              &J012941$-$643447&          &                    &                 &                \\
SGRSC~J0200$-$6007&PMN J0200$-$6007&J020019$-$600541&   ---    & 4.2                &    FR-{\sc ii}   &   ---         \\
                  &                &J020036$-$600825&          &                    &                 &            \\
SGRSC~J0414$-$6933&PMN J0414$-$6933&J041404$-$693407& 0.228:   & 4.5                &   FR-{\sc ii}         & 0.98       \\
               &              &J041358$-$693552&          &                    &                 &            \\
               &              &J041414$-$693159&          &                    &                 &            \\
SGRSC~J1959$-$6402&PKS B1955$-$641   & J195941$-$640420 & --- & 4.9	&  FR-{\sc ii} & --- \\
		  &PMN J1959$-$6403 & J200007$-$640044 &      & 	&	 &      \\
		  &PMN J2000$-$6400 & 		       &      &         &	 &      \\
SGRSC~J2228$-$5600&PMN J2228$-$5559 &J222753$-$560030& 0.0796   & 4.8                &   FR-{\sc ii}         & 0.43     \\
               &              &J222823$-$555953&          &                    &                 &             \\
               &              &J222805$-$555950&          &                    &                 &             \\
SGRSC~J2253$-$5813&PMN J2253$-$5812&J225350$-$581319& ---   & 4.3                &   FR-{\sc ii}         & ---   \\
               &              &J225411$-$581225&          &                    &                 &             \\
SGRSC~J2336$-$8151&PMN J2336$-$8151&J233607$-$815045& 0.1190   & 5.4                &   FR-{\sc ii}         & 0.68   \\
               &              &J233816$-$815332&          &                    &                 &             \\
               &              &J233639$-$815139&          &                    &                 &             \\
\enddata
\end{deluxetable}

\clearpage

\begin{deluxetable}{lcccl}
\tablewidth{0pt}
\tablenum{A3}
\tablecaption{Results of ATCA observations of the sources in Table~A2}
\startdata
\tableline
\tableline
Name  & Core positions    & \multicolumn{2}{c}{Flux density} \\
      & RA~\&~DEC~(J2000~epoch) & 1.4~GHz~(mJy) & 4.8~GHz~(mJy) \\
\tableline
SGRSC~J0020$-$7321& 00 20 43.3 $-$73 21 26.9 & ---    & 2.5 \\
SGRSC~J0129$-$6433& 01 29 26.5 $-$64 33 36.0 & 3.9   &         \\
SGRSC~J0200$-$6007& 02 00 27.4 $-$60 07 04.9 & 2.3   &         \\
SGRSC~J0414$-$6933& 04 14 05.0 $-$69 34 11.5 & 12.9  &         \\
SGRSC~J1959$-$6402& 19 59 53.9 $-$64 02 36.1 & 4.3   &      \\
SGRSC~J2228$-$5600& 22 28 05.7 $-$55 59 48.2 & 40.2  &         \\
SGRSC~J2253$-$5813& 22 53 59.0 $-$58 12 49.8 & 5.4   &         \\
SGRSC~J2336$-$8151& 23 36 39.6 $-$81 51 41.2 & 9.4   &         \\ 
\enddata
\end{deluxetable}

\clearpage

\begin{figure}
(a) \hfill\break
\epsscale{0.3}
\plotone{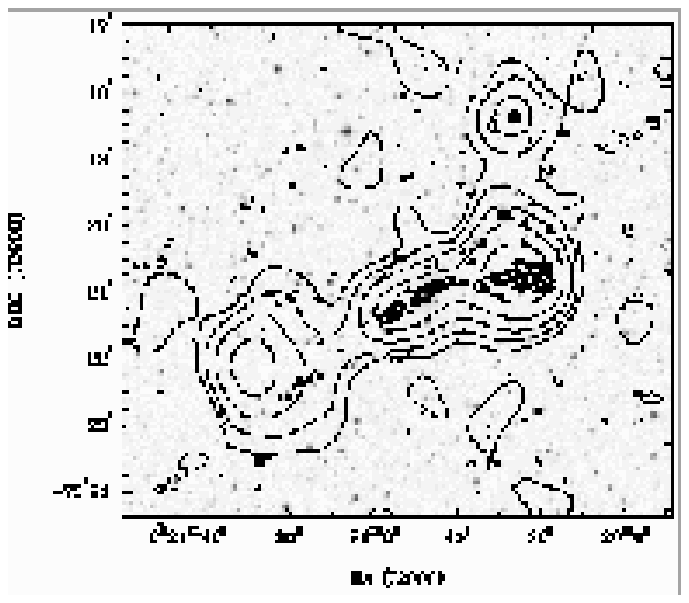}
(b) \hfill\break
\plotone{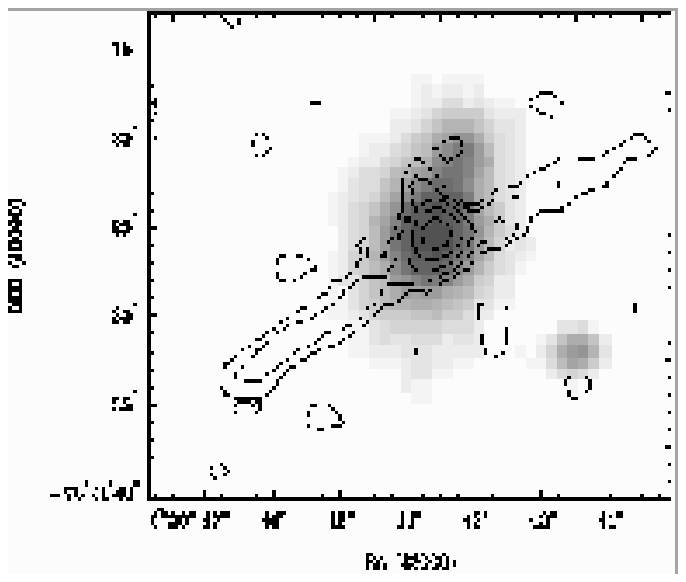}
\epsscale{0.3}
\figurenum{20}
\caption{SGRSC~J0020$-$7321 radio contours on red SSS optical grey scale image. 
ATCA 4.8-GHz image of the core is shown in the lower panel. 
SUMSS contours (thick lines) are drawn at ($-$1, 1, 2, 4, 8, 16) $\times$ 3~mJy~beam$^{-1}$, 
ATCA contours (thin lines) are at ($-$1, 1, 2, 3, 4, 6, 8) $\times$ 0.5~mJy~beam$^{-1}$ (1.4~GHz) and at
($-$1, 1, 2, 3, 4, 6) $\times$ 0.1~mJy~beam$^{-1}$ (4.8~GHz). The beams have FWHM 
$47\farcs 3$~$\times$~$45\arcsec$ (SUMSS), $7\farcs 3$~$\times$~$5\farcs 2$ (ATCA, 1.4~GHz)
and $2\farcs 1$~$\times$~$1\farcs 9$ (ATCA, 4.8~GHz).}
\end{figure}

\begin{figure}
\epsscale{0.3}
\plotone{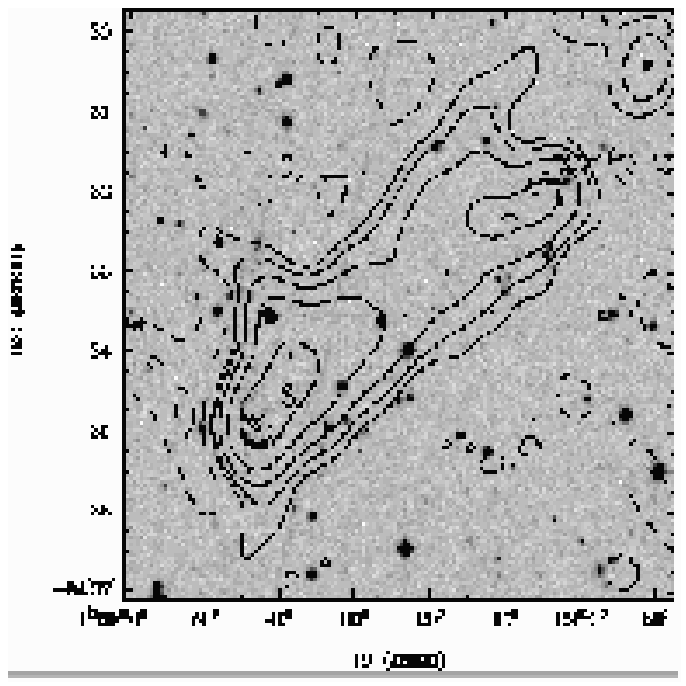}
\figurenum{21}
\caption{SGRSC~J0129$-$6433 radio contours on blue SSS optical grey scale image. 
SUMSS contours (thick lines) are drawn at  
($-$1, 1, 2, 4, 8, 16) $\times$ 3~mJy~beam$^{-1}$ and ATCA contours (thin lines) are at
(1, 2, 3, 4, 6) $\times$ 0.8~mJy~beam$^{-1}$. The beams have FWHM 
$50\farcs 1$~$\times$~$45\arcsec$ (SUMSS) and $6\farcs 5$~$\times$~$5\farcs 8$ (ATCA).}
\end{figure}

\clearpage

\begin{figure}
\epsscale{0.3}
\plotone{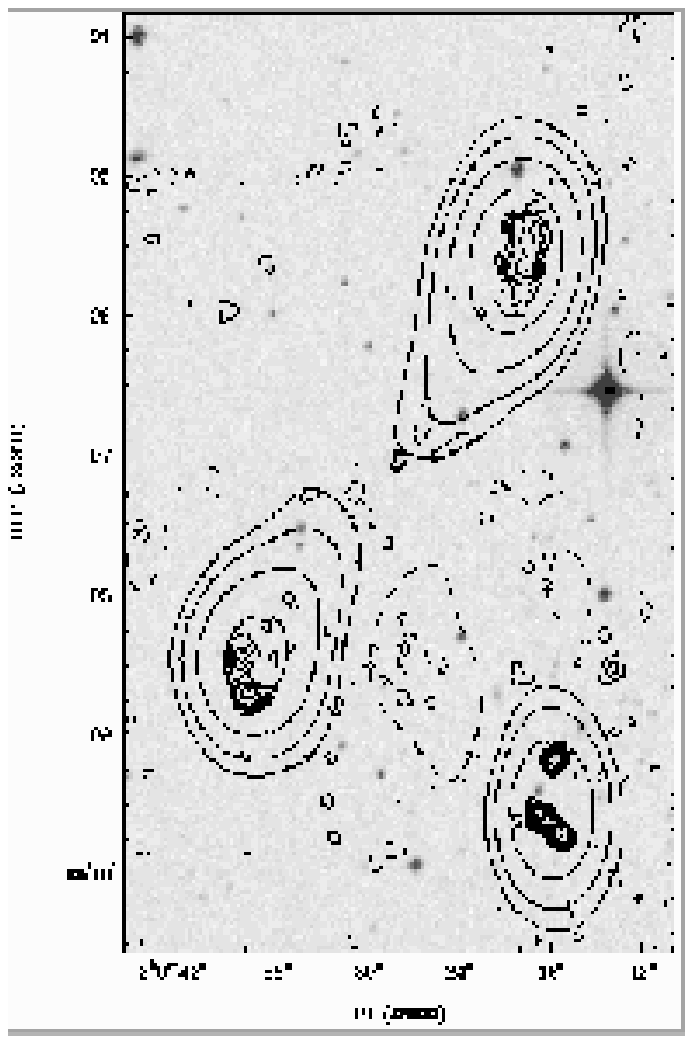}
\epsscale{0.3}
\figurenum{22}
\caption{SGRSC~J0200$-$6007 radio contours on blue SSS optical grey scale image. 
SUMSS contours (thick lines) are drawn at 
($-$1, 1, 2, 4, 8) $\times$ 5~mJy~beam$^{-1}$ 
and ATCA contours (thin lines) are at ($-$1, 1, 2, 3, 4, 6, 8, 12) $\times$ 0.5~mJy~beam$^{-1}$. 
The beams have FWHM 
$52\arcsec$~$\times$~$45\arcsec$ (SUMSS) and $6\farcs 6$~$\times$~$5\farcs 9$ (ATCA).}
\end{figure}

\begin{figure}
(a) \hfill\break
\epsscale{0.1}
\includegraphics[width=40mm]{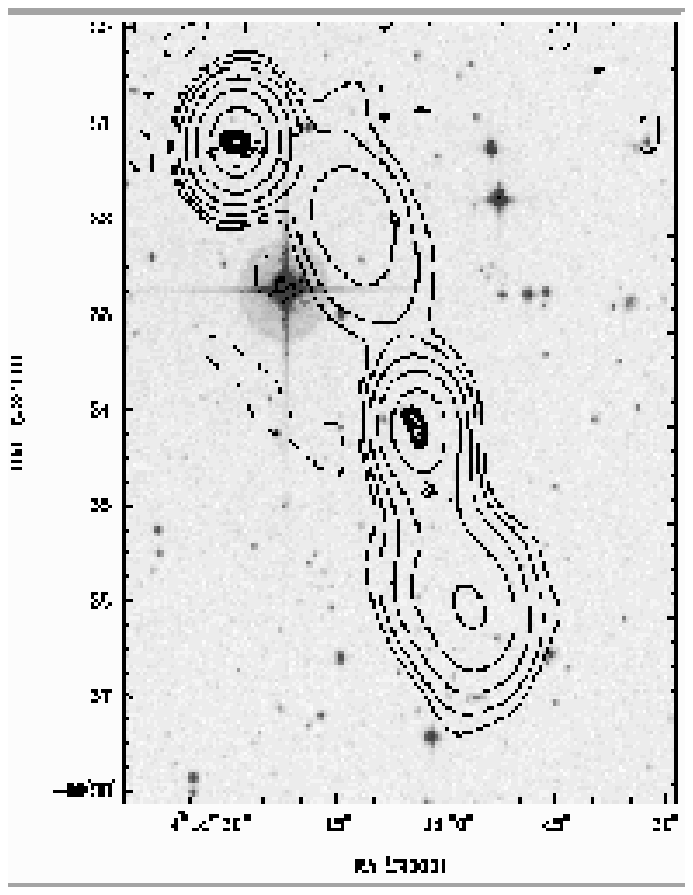} \hfill\break
(b) \hfill\break
\epsscale{0.1}
\includegraphics[width=40mm]{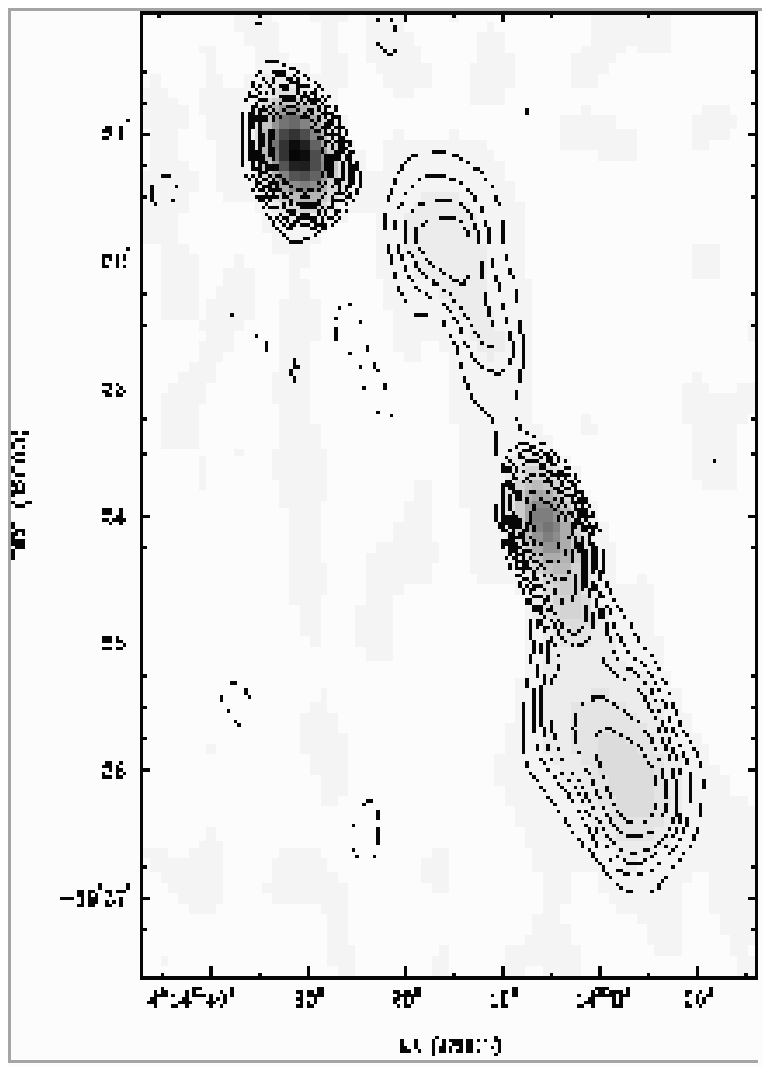}
\figurenum{23}
\caption{SGRSC~J0414$-$6933 radio contours on blue SSS optical grey scale image. 
SUMSS contours (thick lines) are drawn at  
($-$1, 1, 2, 4, 8, 16, 32) $\times$ 3~mJy~beam$^{-1}$, ATCA contours (thin lines) are at 
($-$1, 1, 2, 3, 4, 6, 8, 12, 16) $\times$ 3~mJy~beam$^{-1}$ (panel(a)) and 
($-$2, $-$1, 1, 2, 3, 4, 6, 8, 12, 16, 24, 32) $\times$ 1.2~mJy~beam$^{-1}$ (panel (b)).
The beams have FWHM 
$48\farcs 5$~$\times$~$45\arcsec$ (SUMSS) and $9\arcsec $~$\times$~$6\farcs 2$ (ATCA; panel (a)) and
$32\farcs 7$~$\times$~$21\farcs 5$ (ATCA; panel (b)).}
\end{figure}

\clearpage

\begin{figure}
(a) \hfill\break
\epsscale{0.3}
\plotone{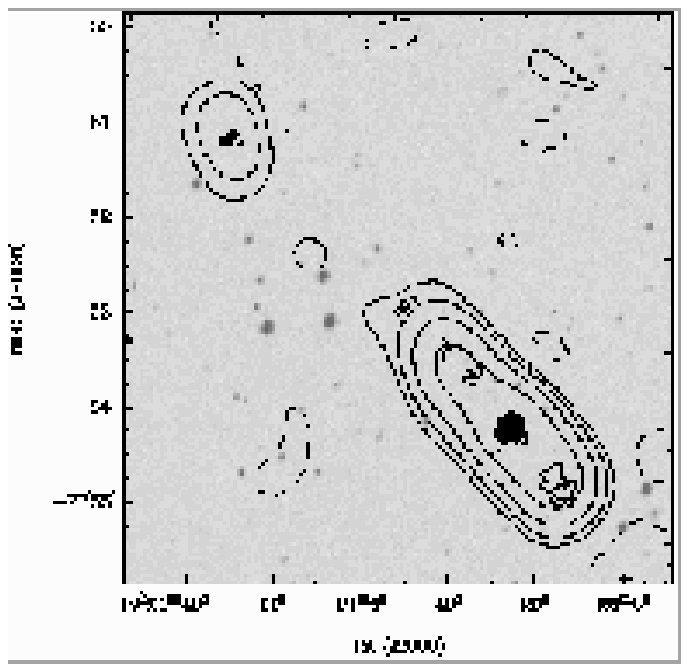}
(b) \hfill\break
\plotone{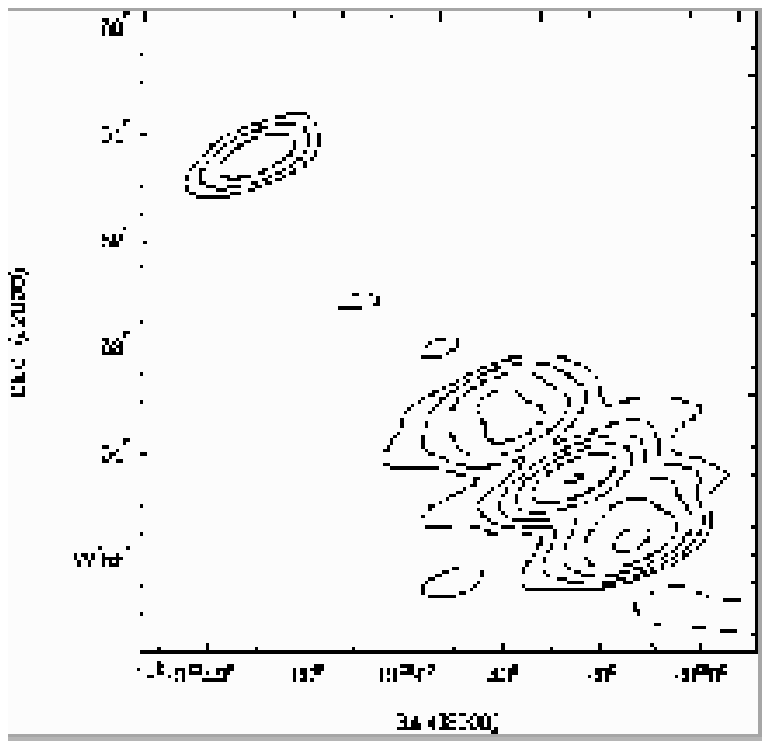}
\epsscale{0.3}
\figurenum{24}
\caption{SGRSC~J1920$-$7753  radio contours on red SSS optical grey scale image. 
SUMSS contours (thick lines) are drawn at ($-$1, 1, 2, 4, 8, 16) $\times$ 3~mJy~beam$^{-1}$ and
ATCA contours (thin lines) are at
($-$1, 1, 2, 3, 4, 6, 8, 12, 16) $\times$ 0.8~mJy~beam$^{-1}$. The beams have FWHM 
$46\farcs 4$~$\times$~$45\arcsec$ (SUMSS) and  $10\farcs 3$~$\times$~$4\farcs 3$ (ATCA). 
A 1.4-GHz ATCA image with a resolution of $52\farcs 2 \times 19\arcsec$ is shown in the lower panel; 
contour levels are at ($-$1, 1, 2, 3, 4, 6, 8) $\times$ 0.8~mJy~beam$^{-1}$.} 
\end{figure}

\begin{figure}
\epsscale{0.3}
\plotone{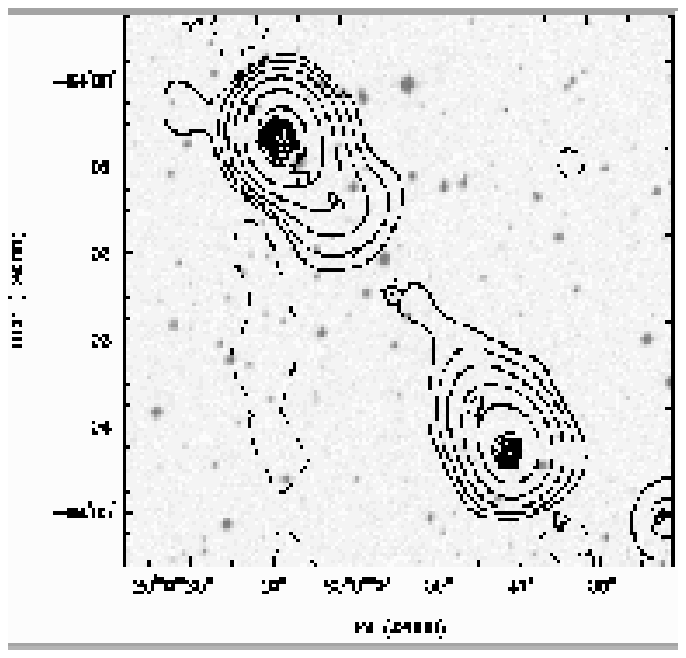}
\figurenum{25}
\caption{SGRSC~J1959$-$6402 radio contours on red SSS optical grey scale image. 
SUMSS contours (thick lines) are drawn at 
($-$1, 1, 2, 4, 8, 16, 32) $\times$ 5~mJy~beam$^{-1}$ and ATCA contours (thin lines) are at
($-$1, 1, 2, 3, 4, 6, 8, 12, 16, 24) $\times$ 3~mJy~beam$^{-1}$. The beams have FWHM 
$50\farcs 1$~$\times$~$45\arcsec$ (SUMSS) and $12\farcs 2$~$\times$~$5\farcs 4$ (ATCA).}
\end{figure}

\clearpage

\begin{figure}
\epsscale{0.3}
\plotone{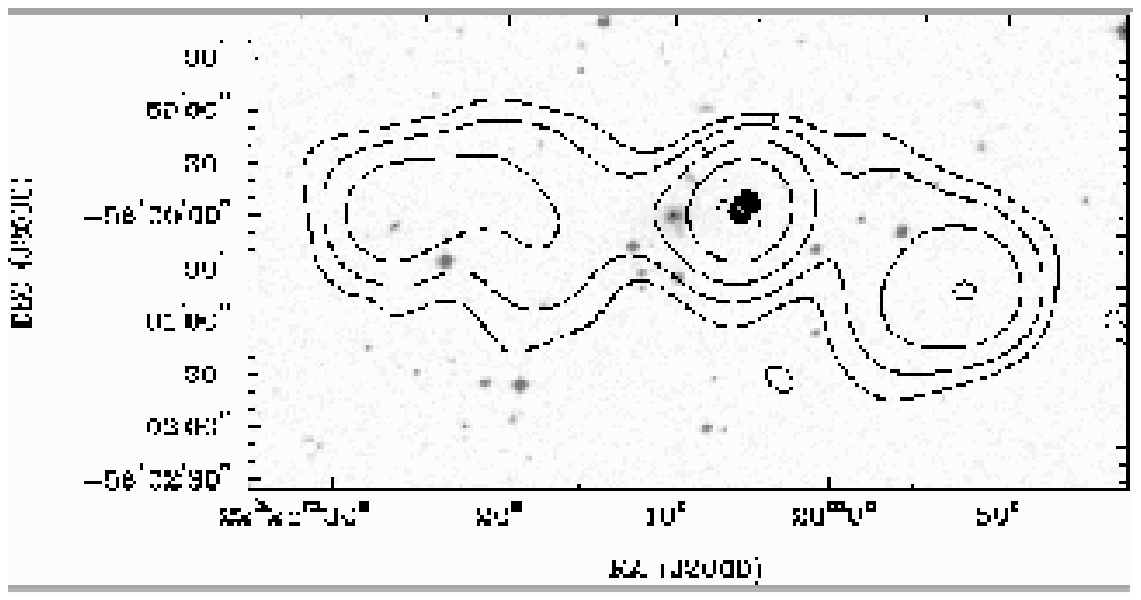}
\figurenum{26}
\caption{SGRSC~J2228$-$5600 radio contours on blue SSS optical grey scale image. 
SUMSS contours (thick lines) are drawn at  
($-$1, 1, 2, 4, 8, 16) $\times$ 3~mJy~beam$^{-1}$ and ATCA contours (thin lines) are at 
($-$1, 1, 2, 3, 4, 6, 8, 12) $\times$ 3~mJy~beam$^{-1}$. The beams used have a FWHM 
$54\farcs 3$~$\times$~$45\arcsec$ (SUMSS) and  $9\farcs 3$~$\times$~$6\farcs 6$ (ATCA).}
\end{figure}

\begin{figure}
\epsscale{0.3}
\plotone{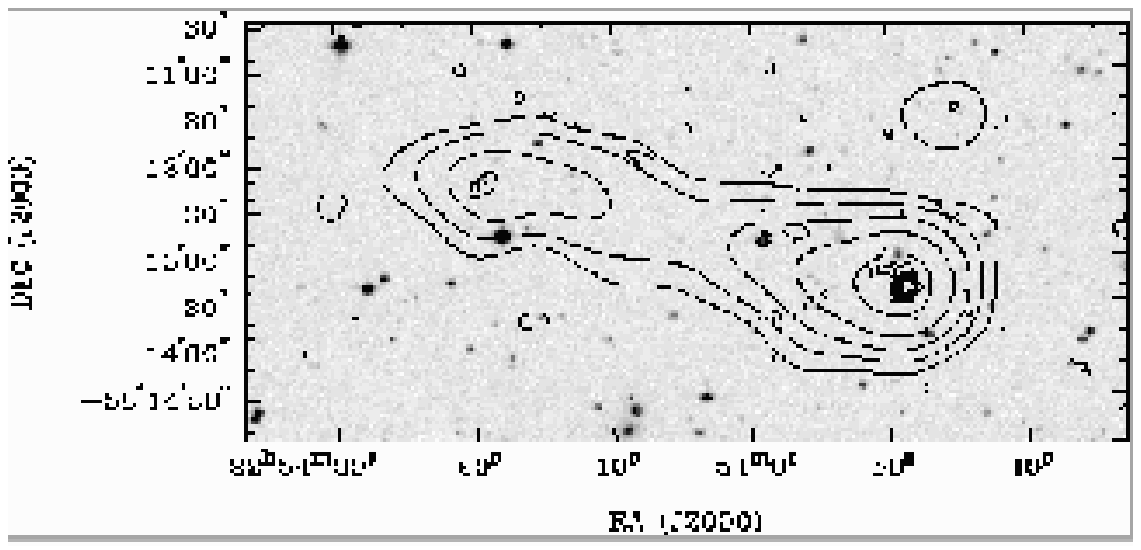}
\figurenum{27}
\caption{SGRSC~J2253$-$5813 radio contours on blue SSS optical grey scale image. 
SUMSS contours (thick lines) are drawn at  
($-$1, 1, 2, 4, 8, 16) $\times$ 4~mJy~beam$^{-1}$ and ATCA contours (thin lines) are at
($-$1, 1, 2, 3, 4, 6, 8, 12) $\times$ 2~mJy~beam$^{-1}$. The beams have FWHM 
$52\arcsec $~$\times$~$45\arcsec$ (SUMSS) and  $8\farcs 9$~$\times$~$6\farcs 6$ (ATCA).}
\end{figure}

\begin{figure}
\epsscale{0.3}
\plotone{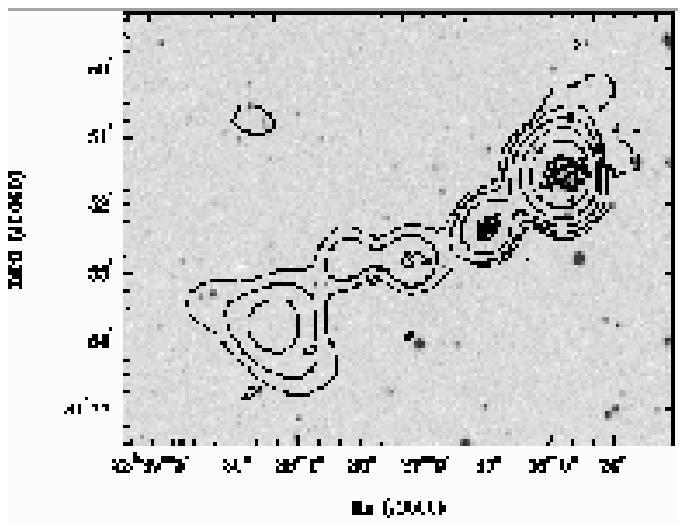}
\figurenum{28}
\caption{SGRSC~J2336$-$8151 radio contours on blue SSS optical grey scale image. 
SUMSS contours (thick lines) are drawn at 
($-$2, $-$1, 1, 2, 4, 8, 16, 32) $\times$ 4~mJy~beam$^{-1}$ and ATCA contours (thin lines) are at 
($-$1, 1, 2, 3, 4) $\times$ 2.5~mJy~beam$^{-1}$. The beams have FWHM 
$45\farcs 7 $~$\times$~$45\arcsec$ (SUMSS) and $8\farcs 8$~$\times$~$5\farcs 9$ (ATCA).}
\end{figure}

\clearpage

\begin{figure*}
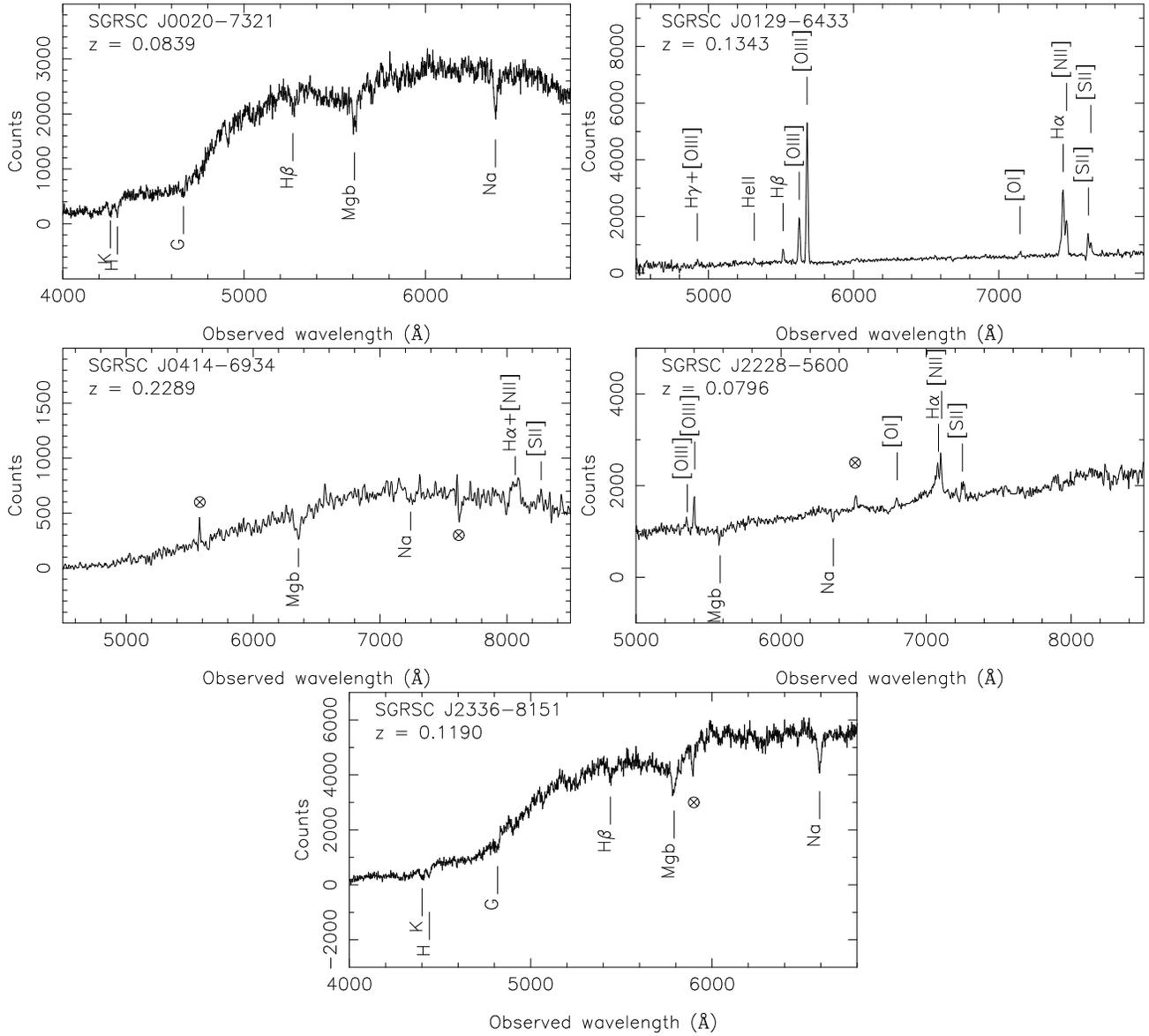

\centerline{
\includegraphics[height=85mm,angle=-90]{f29a.eps}
\includegraphics[height=85mm,angle=-90]{f29b.eps}
}
\centerline{
\includegraphics[height=85mm,angle=-90]{f29c.eps}
\includegraphics[height=85mm,angle=-90]{f29d.eps}
}
\centerline{
\includegraphics[height=85mm,angle=-90]{f29e.eps}
}
\figurenum{29}
\caption{
Optical spectra of the candidates that are not in the complete 
sample of giant radio sources.
}
\end{figure*}

\end{document}